\documentclass[fleqn,usenatbib]{mnras}
\usepackage{newtxtext,newtxmath}
\usepackage[T1]{fontenc}
\usepackage{enumerate}

\usepackage{graphicx}
\usepackage{amsmath}

\usepackage{color}
\definecolor{crevision}{rgb}{0.8, 0.2, 0.8}

\defcitealias{lambrechts2014a}{L2014}
\defcitealias{bitsch2018}{B2018}

\title[multi-species pebble accretion]{Planetary core formation via multi-species pebble accretion}

\author[Andama et al.]{
G. Andama,$^{1}$\thanks{E-mail: gandama@must.ac.ug}
N. Ndugu,$^{2,3}$
S.K. Anguma,$^{3}$
E. Jurua$^{1}$
\\
$^{1}$Department of Physics, Mbarara University of Science and Technology, Mbarara, Uganda\\
$^{2}$Center for Space Research, North-West University, South Africa\\
$^{3}$Department of Physics, Muni University, Arua, Uganda\\
}

\date{Accepted XXX. Received YYY; in original form ZZZ}

\pubyear{2021}

\begin{document}
\label{firstpage}
\pagerange{\pageref{firstpage}--\pageref{lastpage}}
\maketitle
\numberwithin{equation}{section}
\numberwithin{figure}{section}
\begin{abstract}
In the general classical picture of pebble-based core growth, planetary cores grow by accretion of single pebble species. The growing planet may reach the so-called pebble isolation mass, at which it induces a pressure bump  that blocks inward drifting pebbles exterior to its orbit, thereby stalling core growth by pebble accretion. 
In recent hydrodynamic simulations, pebble filtration by the pressure bump depends on several parameters including core mass, disc structure, turbulent viscosity and pebble size. We investigated how accretion of multiple, instead of single, pebble species affects core growth rates, and how the dependence of pebble isolation mass on turbulent viscosity and pebble size sets the final core masses. We performed numerical simulations in viscous 1D disc, where maximal grain sizes were regulated by grain growth, fragmentation and drift limits. We confirm that core growth rates and final core masses are sensitive to three key parameters: threshold  velocity at which pebbles fragment on collision, turbulent viscosity and distribution of pebble species, which yield a diversity of planetary cores. With accretion of multiple pebble species, planetary cores can grow pretty fast, reaching over 30 -- 40 $M_{\rm{E}}$ in mass. Potential cores of cold gas giants were able to form from embryos initially implanted as far as 50 AU. Our results suggest that accretion of multi-species pebbles could explain: the estimated 25 -- 45 $M_{\rm{E}}$ heavy elements abundance inside Jupiter's core; massive cores of extrasolar planets; disc rings and gaps at wider orbits; early and rapid formation of planetary bodies.
\end{abstract}




\section{Introduction}
The theory of planet formation in protoplanetary discs (hereafter, PPDs) has developed by leaps and bounds ever since the monumental work of \citet{safronov1969}. In one school of thought, planets may form by gravitational collapse of dense and dynamically cold gas disc~\citep{kuiper1951, cameron1978, boss1997,gammie2001,rice2003,tanga2004,rafikov2005,durisen2006} followed by tidal downsizing~\citep{nayakshin2010}. Gravitational collapse requires sufficiently massive discs and mainly favours the formation of giant planets at the disc outskirts~\citep[][]{boss1997,boley2009,armitage2010}. 

In another school of thought, planets may also form oligarchically by core accretion paradigm~\citep[][]{wetherill1980,kokubo1998,thommes2003,coleman2014}. Here, micrometre-sized dust grains in the natal PPDs first have to grow by coagulation into millimetre-centimetre (mm -- cm) sized particles. These mm -- cm sized particle may concentrate in some regions of the disc where they may gravitationally collapse into metre-kilometre sized bodies called planetesimals \citep[e.g.,][]{youdin2005,johansen2007b,raettig2015,carrera2015}. Planetesimals larger than 100 km then form potential planetary embryos.  These planetary embryos can also form through a different mechanism, for example, when collisions between smaller planetesimals result into a merger of over 100 km sized planetesimals~\citep[e.g.,][]{kokubo2012}. In the core accretion paradigm, the planetary embryos can then accrete smaller planetesimals to grow into a full planet~\citep[][]{safronov1969,mizuno1978,mizuno1980,kokubo2012}. However, core growth via planetesimal accretion is typically slow unless most of the  solid mass in disc is converted into planetesimals less than 10 km in size~\citep[][]{tanaka1999,thommes2003,levison2010,johansen2019b}. Nevertheless, fast planetesimal-based core accretion rates were reported for planetesimals with size less than 1 km~\citep[e.g.,][]{mordasini2009,mordasini2009b}, even though there are no evidences in the solar system for planetesimals of such smaller sizes~\citep[][]{bottke2005,bottke2005b, morbidelli2009,singer2019}. 

A planetary core may also grow into a full planet by accreting aerodynamically coupled bodies via gas drag, popularly known as pebble accretion~\citep{johansen2010, ormel2010,lambrechts2012,lambrechts2014a}. Though the current framework of planet formation by core accretion of planetesimals or pebbles is the most successful, it cannot satisfactorily attribute the observed substructures at wider orbital locations to planets~\citep{lodato2019,ndugu2019,nayakshin2019}.

The formation of planetary bodies by core accretion paradigm is strongly shaped by the availability and size distribution of solid material in the PPDs~\citep[see review by][]{johansen2014}. In particular, pebble accretion is constrained by the formidable radial drift barrier which leads to rapid loss of centimetre-sized solids on short dynamical time-scales~\citep{whipple1972,weidenschilling1977,takeuchi2005, alexander2007,brauer2007, brauer2008}. Furthermore, \citet{johansen2019} demonstrated that mm-cm sized dust material with stokes number larger than 0.1 are expected to drain on shorter time scales than the disc life time. Consequently such rapid loss of dust grains can impede formation of planetesimals and the subsequent growth of planetary cores by pebble accretion. 

However, studies of PPDs from various surveys~\citep[e.g.,][]{testi2003,wilner2005,rodmann2006,brauer2007,perez2012,trotta2013,carrasco2016,ansdell2017} indicate that substantial amount of dust grains in the mm -- cm range survive even in the discs that are in their late stages of evolution, contrary to theoretical predictions. The findings from the above surveys may not be globally true. In fact, recent study by~\citet{tychoniec2020} revealed that most discs contain small amount of pebbles and only few discs retain substantial amount of pebbles. The retention of small dust grains in PPDs has been linked to destructive collision of larger dust aggregates~\citep{blum2008} and subsequent coagulation-fragmentation equilibrium~\citep{dominik2008}. \cite{birnstiel2009} demonstrated that fragmentation of grains could facilitate dust retention in the disc.  Also zonal flows formed by magnetorotational instability (MRI)~\citep[e.g.,][]{johansen2009,dzyurkevich2010,johansen2011,uribe2011} can cause over-densities and hence pressure bumps that act as dust traps and help to retain dust grains in the outer region of the disc~\citep{pinilla2012}. Dust trapped in the pressure bumps induced by massive planets may also undergo fragmentation to produce finer grains~\citep{drazkowska2019}. The retention, evolution and distribution of grain sizes play an important role in planet formation models as they determine the outcome of the planetary bodies~\citep{fouchet2005}, as well as the disc structure~\citep{dullemond2004}.

The grain size distribution can be approximated using either the simple power-law fits as in the MRN model~\citep[][]{mathis1977} or the complex analytic approach of ~\cite{birnstiel2011}.~\cite{birnstiel2012} further developed a two-dust population model of dust size distribution, classifying the distribution into opacity bearing grains and larger grains. The subpopulation of the smaller grains determines the temperature profile and hence the disc structure~\citep{dullemond2004, savvidou2020}.

The discussed observational surveys of dust disc and dust distribution simulations present piece of evidence for the existence of numerous dust species in PPDs, each with unique spatio-temporal distribution of Stokes number. The assortment of different grain sizes may point to the fact that core accretion proceeds by accretion of multiple pebble species. 

However, previous works that studied core growth via pebble accretion ~\citep[e.g.,][]{lambrechts2014a,lambrechts2014b, guillot2014, morbidelli2015, bitsch2015b, bitsch2017, ndugu2017,brugger2018,johansen2019, ndugu2019,ndugu2021} were based on two standard prescriptions.
Firstly, the studies used a single spatio-temporal dominant particle size or Stokes number, which is assumed to carry most of the solid mass. Dust coagulation models \citep[e.g.,][]{dullemond2005} predict rapid conversion of most of the dust  into larger grains within very short time scales compared to the disc lifetimes.  The grain population can thus be modeled as consisting of small- and large-size grain populations, where most of the mass is carried by the large-size population~\citep{birnstiel2012}. Through coagulation-fragmentation equilibrium, grain sizes may attain quasi-stationary size distribution \citep{dullemond2005}. A quasi-stationary size means that one is justified to use single size approach, especially when the resulting grain sizes in the distribution are very similar. 
Nevertheless, in this work, instead of a single pebble size or Stokes number, we have studied how grain size distributions reconstructed from the dust evolution model of~\cite{birnstiel2012} influences planetary core growth through accretion of different pebble species. This is because considering only dominant pebble species may not provide a complete picture of the final core masses and growth rates. This could be important, especially when several other pebble species in the distribution carry some considerable mass. Consequently, we  can easily underestimate final core masses or growth times if some of the species that may also contain significant mass are neglected during core assembly, as may be the case in single species approach.

 Secondly, core growth stops whenever a pressure bump is induced by the growing core, and hence the final core masses are fixed by the classical pebble isolation mass~\citep[][hereafter L2014]{lambrechts2014a}. However, hydrodynamical simulations show that through turbulence, some pebble species can still diffuse through the pressure bump exerted by sufficiently massive cores and can even pass into gaps carved out by Jupiter-mass planets~\citep{weber2018}. As a result of turbulent diffusion, ~\cite{bitsch2018}~(hereafter, B2018) and~\cite{ataiee2018} further demonstrated that pebble isolation mass might not be completely universal for all pebble species since smaller pebble species may overcome weaker pressure bumps. This therefore suggests that those pebble species that diffuse through the pressure bump may sustain core accretion for an extended period.

The grain size distribution as in the two-population model of \cite{birnstiel2012} now  provides an opportunity to study core accretion in the context of multiple dust species~\citep{guilera2020, venturini2020b, drazkowska2021,savvidou2021, schneider2021}. For instance, in both \cite{guilera2020}  and \cite{venturini2020b}, the authors used a mass weighted representative pebble size derived from dust population of several species in their pebble accretion model, where the final core masses are fixed by classical \citetalias{lambrechts2014a} prescription. 
Using full grain size distribution, \cite{guilera2020} studied how giant planet cores can form by hybrid pebble and planetesimal accretion at pressure maxima. In~\cite{venturini2020b}, the authors focused on the formation of super -- Earths inside the snow line. On the other hand, \cite{schneider2021} studied the heavy element content of giant planets, also based on full dust evolution, where they focused on evolution of different chemical species rather than the physical size distributions as considered in this work.

In~\citet{drazkowska2021}, using full size distribution, the authors studied the impact of grain growth and fragmentation on core growth rate while the planet accreted throughout at a particular radial location. This allowed the authors to exclusively study the actual impact of fragmentation on core growth. Hence they performed a limited set of simulations and did not take into account many aspects of planet formation such as gas accretion and orbital migration, which are crucial for the final architecture of planetary systems. For instance, inward migration could enable a core to reach pebble isolation mass much faster in the disc regions closer to the star where the isolation mass is lower. This then gives the planet a chance to accrete gas and grow into a gas giant.  Lastly, in their work the core masses were measured based on the \citetalias{lambrechts2014a} model without taking into account the diffusion of pebbles across the pressure bump, which may impact the final core mass as discussed in \citetalias{bitsch2018}.

In this study, we performed similar numerical simulations, but quite different from the one presented in~\citet{drazkowska2021}. The major difference in our work is that we self-consistently reconstruct a distribution of pebble sizes, their corresponding mass and Stokes numbers after full dust evolution at every time step during core growth. This allowed us to investigate the contribution of each individual grain species as opposed to the mass averaged pebble flux model used in previous studies. Here we focused on two major problems: how concurrent accretion of several pebble species may impact core growth; how the dependence of pebble isolation mass on turbulent viscosity and pebble sizes may determine final core masses. All this is based on the two-population model of~\cite{birnstiel2012} and the size reconstruction recipe in~\cite{birnstiel2015}.  We did not perform hydrodynamic simulations of pebble isolation mass because it is beyond the scope of this work but rather used the formula from~\citetalias{bitsch2018} to calculate the final masses.

The rest of the paper is structured as follows: In Section~\ref{theory}, we describe the underlying disc model, size distribution of particles and the core growth model. In Section~\ref{simulations}, we explain the main numerical experiments that we performed. We present and discuss our results in Section~\ref{results}. We then summarise our findings in Section~\ref{conclusions}.

\section{Theoretical background}\label{theory}
\subsection{The disc evolution model}
In order to provide a complete picture of our core accretion of multi-pebble species model, we employed the two-population model of~\cite{birnstiel2012}. Here, we only describe the key ingredients of the gas and dust evolution model that we adopted in our numerical simulations and we refer the reader to~\cite{birnstiel2009, birnstiel2010, birnstiel2011, birnstiel2012} for more detailed description of dust size evolution.

In the simulations, the initial gas surface density $\Sigma_{\rm{g}}$ is calculated using the self-similar solution of~\cite{lyndenbell1974} given by
\begin{equation}
\Sigma_{\rm{g}}(r) \propto \left( \frac{r}{r_{\rm{c}}} \right)^{-\gamma}{\rm{exp}}\left[-\left( \frac{r}{r_{\rm{c}}} \right)^{2-\gamma} \right].
\end{equation}
Here, $\gamma$ is the viscosity power-law index and $r_{\rm{c}}$ the characteristic radius at an initial time $t_{0}$.

For dust evolution in the disc, we adopt the two dust population model of \cite{birnstiel2012}, where the dust surface density evolves according to the advection-diffusion equation:
\begin{equation}
 \frac{\partial\Sigma_{\rm{p}}}{\partial t}+\frac{1}{r}\frac{\partial}{\partial r}\left[r\left(\Sigma_{\rm{p}}\bar{u}-D_{\rm{g}}\Sigma_{\rm{g}}\frac{\partial}{\partial r}\left(\frac{\Sigma_{\rm{p}}}{\Sigma_{\rm{g}}} \right) \right) \right]=0. \label{eq:advection}
\end{equation}
Here, $\Sigma_{\rm{p}}$ is the dust and gas surface densities and $D_{\rm{g}}$ is the gas diffusivity. $\bar{u}$ is the velocity of the dust weighted by the mass of the two dust populations and is given by
\begin{equation}
 \bar{u}=(1-f_{\rm{m}})u_{0}+f_{\rm{m}}u_{1}, \label{eq:dustvelocity}
\end{equation}
where $u_{0}$ and $u_{1}$ are respectively radial velocities of the two small- and large-size grain populations, with representative sizes $a_{0}$ and $a_{1}$ which are set by growth, drift and fragmentation limits as described in ~\cite{birnstiel2012}. Here,  the mass fraction $f_{\rm{m}}$ of the large-size grain population at radial distance $r$ is calibrated  as 0.97  and 0.75 for drift and fragmentation limited regimes, respectively.

The radial velocities $u_{0}$ and $u_{1}$ are calculated as the sum of radial drift velocity and the radial velocity due to gas drag~\citep{weidenschilling1977}:
\begin{equation}
 u_{i}=-\frac{2\tau_{\rm{i}}}{1+\tau_{\rm{i}}^{2}}u_{\eta}+\frac{1}{1+\tau_{\rm{i}}^{2}}u_{\rm{g}},
\end{equation}
 where $u_{\rm{g}}$ is the gas velocity, $u_{\rm{i}}~(i=0,1)$ represents either $u_{0}$ or $u_{1}$, and  $\tau_{\rm{i}}$ the corresponding Stokes numbers, which will be discussed in Section~\ref{sec:particledistro}. Here $u_{\eta}$ is the headwind velocity~\citep{weidenschilling1977,nakagawa1986} given by
\begin{equation}
 u_{\eta}=-\frac{1}{2} \frac{1}{\rho_{\rm{g}}\Omega_{\rm{K}}}\frac{\partial~P}{\partial~r}.
\end{equation}

where $\rho_{\rm{g}}$ is the mid-plane gas density, $P$ the gas pressure and the $\Omega_{\rm{K}}$ the Keplerian frequency. The gas velocity $u_{\rm{g}}$ is given by
 \begin{equation}
  u_{\rm{g}}=c_{\rm{s}}\sqrt{1.5\alpha_{\rm{t}}},
 \end{equation}
where $c_{\rm{s}}$ is the sound speed and $\alpha_{\rm{t}}$ is the turbulence parameter.
\begin{figure}
 \includegraphics[width=0.5\textwidth]{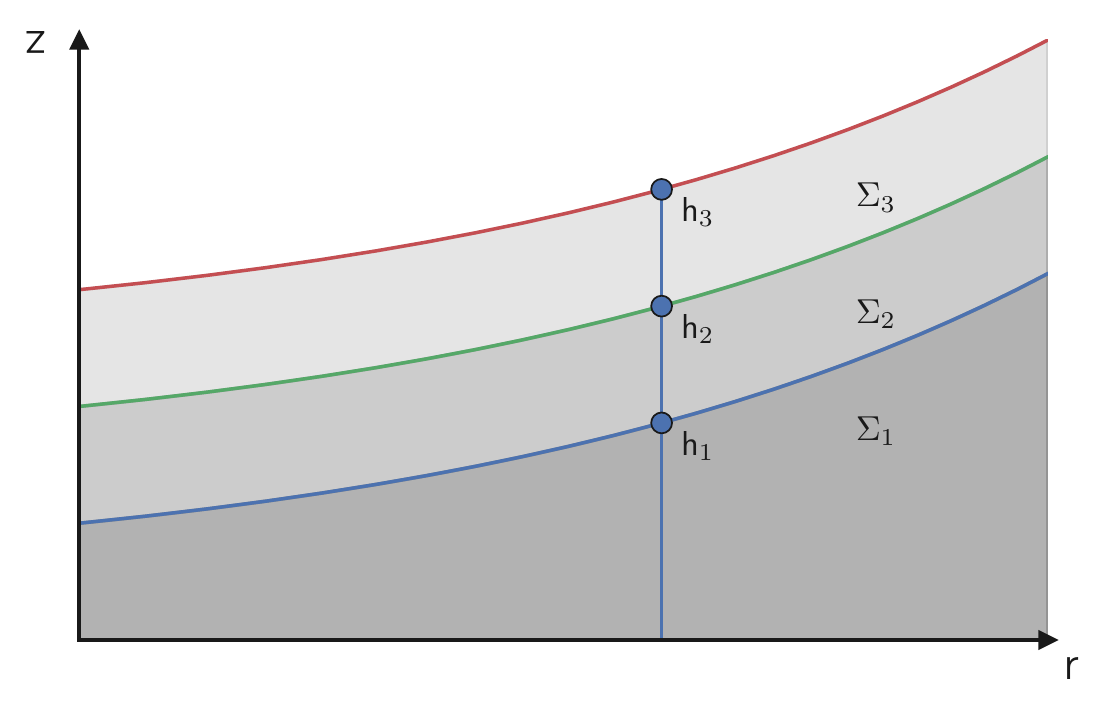}
 \caption{The geometry of vertical particle distribution used in our core growth model, where the particle species change from 1 -- 3 with decreasing size.}
\label{fig:geometry}
\end{figure}

\subsection{Particle size distribution}\label{sec:particledistro}

The distribution of solids in PPDs depends on their aerodynamic properties characterised by  friction time, $t_{\rm{f}}$, given by~\citep{whipple1972, weidenschilling1977} as
\begin{equation}
 t_{\rm{f}}=\frac{m v}{||F_{\rm{D}}||}, \label{eq:Stokes1}
\end{equation}
where $m$ is the mass of the particle, $v$ the terminal velocity and $||F_{\rm{D}}||$ the gas drag force. 

For small particles, it is usually convenient to express their degree of coupling with the gas in terms of the dimensionless Stokes number $\tau_{\rm{s}}$ given by
\begin{equation}
 \tau_{\rm{s}}=t_{\rm{f}}\Omega_{\rm{K}}=\frac{\rho_{\bullet}R}{\rho_{\rm{g}}h_{\rm{g}}}= \frac{\rho_{\bullet}\pi R}{2\Sigma_{g}}, \label{eq:Stokes2}
\end{equation}
 where $\rho_{\bullet}$ is the material density, $R$ the particle size, $\Sigma_{\rm{g}}$ the gas surface density,  $h_{\rm{g}}$ the gas scale height.

Bodies with Stokes number $\tau_{\rm{s}}>1$ become increasingly less coupled to the gas or even decouple completely, and therefore are not well suited for core growth by aerodynamic drag in the settling regime. On the other hand, bodies with Stokes number smaller than 0.001 have short friction time and hence long settling time that enables them to stay coupled to the gas during their gravitational encounter with the growing cores. Therefore, these particles mostly follow the gas streamlines and may not be accreted efficiently~\citep[][]{guillot2014, johansen2019,rosenthal2020}. Furthermore, as shown by~\cite{johansen2019}, pebbles with Stokes number $\sim 0.1$ drift faster and are lost to the central star on short dynamical scales. 

Particles settle towards the mid-plane depending on their size and material density and may also radially drift on time-scales which equally depend on their aerodynamic properties~\citep[][]{whipple1972, adachi1976, weidenschilling1977, dullemond2004}. For higher radial drift velocities, larger particles tend to concentrate more in the inner disc regions than the smaller particles~\citep[][]{fouchet2005,birnstiel2010, birnstiel2011,testi2014}. This results into vertical and radial stratification of particles (see Figure~\ref{fig:geometry}).

In a quiescent disc, small dust grains are found to settle to the mid-plane, while the larger grains tend to oscillate about the mid-plane as their oscillation amplitude decays~\citep[][]{garaud2004a}.  Here, we assume a simple model with N particle species, where particles sediment in a layered fashion with different scale heights~\citep[][]{fouchet2005}. Particles  with small Stokes number stay in the upper layers of the gas disc, and those with large Stokes number settle in the mid-plane as depicted in the schematic shown in Figure~\ref{fig:geometry}. In this picture, the scale height, $h_{\rm{i}}$ (with i=1, 2, ..., N denoting different species) of each particle species is then calculated from~\citep{youdin2007} as
\begin{equation}
 h_{\rm{i}}=\left(\frac{\alpha_{\rm{t}}}{\tau_{\rm{i}}} \right)^{1/2}h_{\rm{g}}, \label{eq:05}
\end{equation}
where  $\tau_{\rm{i}}$ is the Stokes number of a particular species. 

The differential settling of particles discussed above suggests that the upper layers of the gas disc become relatively devoid of dust even for $\mu$m grains~\citep[][]{dullemond2004}. This motivates us to consider N different pebble species with surface densities, $\Sigma_{\rm{p,i}}$. The simple MRN model of~\cite{mathis1977} describes grain size distribution quite well. However, we used the two-population model of~\cite{birnstiel2012} in our work from which we reconstruct the grain sizes and their surface densities, $\Sigma_{\rm{p,i}}$ as presented in~\cite{birnstiel2015}. This is because we want to use a self-consistent size distribution which takes into account the balance between grain growth, fragmentation and drift barrier as described in~\cite{birnstiel2012}. 

 Our grain sizes are distributed such that $R_{\rm{i+1}}=1.12R_{\rm{i}}$ as in~\cite{birnstiel2011} where $R_{\rm{i}}$ is the size of the i-th species. The dust surface density, $\Sigma_{\rm{p,i}}$,  corresponding to each species at a given radial distance and time is then reconstructed from gas surface density ($\Sigma_{\rm{g}}$), dust surface density ($\Sigma_{\rm{p}}$), fragmentation velocity ($\it{u}_{\rm{f}}$), turbulent strength ($\alpha_{\rm{t}}$), material density ($\rho_{\bullet}$) and the disc mid-plane temperature (T) according to the recipe given in~\citet{birnstiel2015}.

Laboratory experiments have constrained the fragmentation velocity threshold for silicate grains to be $\it{u}_{\rm{f}}$ = 1 m/s~\citep{blum2008}, while numerical simulations show that water-ice aggregates which can grow to centimetre sizes tend to fragment at much higher velocities of $\it{u}_{\rm{f}}\geq$ 10 m/s~\citep{brauer2008,wada2008,gundlach2011,gundlach2015}. But recent laboratory experiment by~\cite{musiolik2019} seem to suggest that ice and silicate grains have similar fragmentation velocities.  However, the collisional outcomes also depend on the turbulent strength, the internal density of the solid bodies and the local temperature which can lead to wide range of particle sizes. 
Lastly, the corresponding Stokes number, $\tau_{i}$, of each pebble species from the reconstructed size distribution is then calculated using Equation~(\ref{eq:Stokes2})


\subsection{Dominant size distribution}\label{sec:dominantspecies}
In the two-population model of~\cite{birnstiel2012}, the surface densities $\Sigma_{0}$ and $\Sigma_{1}$ of the small and large populations at a radial distance $r$ are calculated as
\begin{align}
 \Sigma_{0}&=\Sigma_{\rm{p}}(r)(1-f_{\rm{m}}(r)) &\\
 \Sigma_{1}&=\Sigma_{\rm{p}}(r)f_{\rm{m}}(r).
\end{align}

In typical classical pebble accretion scenarios, a single representative size is assumed to contain most of the dust mass, for example, in the large population. In this classical picture we can calculate the mass averaged dominant size $R_{\rm{d}}$ for the reconstructed size distribution as~\citep{guilera2020, venturini2020b}
\begin{equation}
 R_{\rm{d}}=\frac{\sum_{i}\epsilon_{\rm{i}}R_{\rm{i}}}{\sum_{\rm{i}}\epsilon_{\rm{i}}},
\end{equation}
where 
\begin{equation}
 \epsilon_{\rm{i}}=\frac{\Sigma_{\rm{p,i}}}{\Sigma_{\rm{g}}}\left(\frac{\alpha_{\rm{t}}+\tau_{\rm{i}}}{\alpha_{\rm{t}}} \right)^{1/2}.
\end{equation}
The corresponding Stokes number for the dominant species with size $R_{\rm{d}}$ can then be obtained by using Equation~(\ref{eq:Stokes2}).

\subsection{Core growth model} \label{coregrowth}
In our planetary growth model, planetary embryos start accreting pebbles at the transition mass where Hill accretion becomes more efficient than the Bondi regime~\citep{lambrechts2012}.
For the governing equations, we start from the classical pebble accretion rate of a dominant pebble species $i$ in the 2-D Hill regime given by~\citep{morbidelli2015}:
\begin{equation}
 \dot{M}_{\rm{2D}}=\left \{
 \begin{array}{ll}
  2\left(\tau_{\rm{i}}/{0.1} \right)^{2/3}\Omega_{\rm{K}} r_{H}^{2}\Sigma_{\rm{p,i}}& (\tau_{\rm{i}} < 0.1)\\ \\
  2\Omega_{\rm{K}} r_{H}^{2}\Sigma_{\rm{p,i}}&(\tau_{\rm{i}} \ge 0.1)
 \end{array}
 \right. \label{eq:06}
\end{equation}
where $r_{\rm{H}}$ is the Hill radius. 

The 2-D solid accretion takes place when the effective accretion radius is greater than the particle scale height, otherwise the core grows by 3-D accretion. Since the protoplanets are initially small, their gravitational reach is mostly below the particle scale height and hence the accretion rate follows the 3-D mode.  The 3-D and 2-D accretion modes are then related as in~\cite{morbidelli2015} by 
\begin{equation}
 \dot{M}_{\rm{3D}}=\left[\sqrt{\frac{\pi}{8}}\left(\frac{\tau_{\rm{i}}}{0.1}\right)^{1/3}\frac{r_{\rm{H}}}{h_{\rm{i}}}\right]\dot{M}_{\rm{2D}}, \label{eq:07}
\end{equation}
where we can calculate the critical core masses at which accretion switches from 3-D to 2-D as
\begin{equation}
 M_{\rm{3D2D}}=4.06\times10^{5}\times\alpha_{\rm{t}}^{3/2}\tau_{\rm{i}}^{-5/2}\left(\frac{H}{r} \right)^{3}~M_{\rm{E}},\label{eq:CriticalMass}
\end{equation}
Where $H/r$ is the disc aspect ratio. Hence the core accretion rate of the $i{\rm{-th}}$ pebble species is given by 
\begin{equation}
 \dot{M}_{\rm{core,i}}=\left \{
 \begin{array}{ll}
  \dot{M}_{\rm{2D}} & \rm{for~}~\sqrt{\frac{\pi}{8}}\left(\frac{\tau_{\rm{i}}}{0.1} \right)^{1/3}r_{\rm{H}}>h_{\rm{i}}\\
  \dot{M}_{\rm{3D}}&~\text{otherwise}
 \end{array}
 \right. \label{eq:CoreGrowthRegime}
\end{equation}

\begin{figure*}
\includegraphics[width=\textwidth]{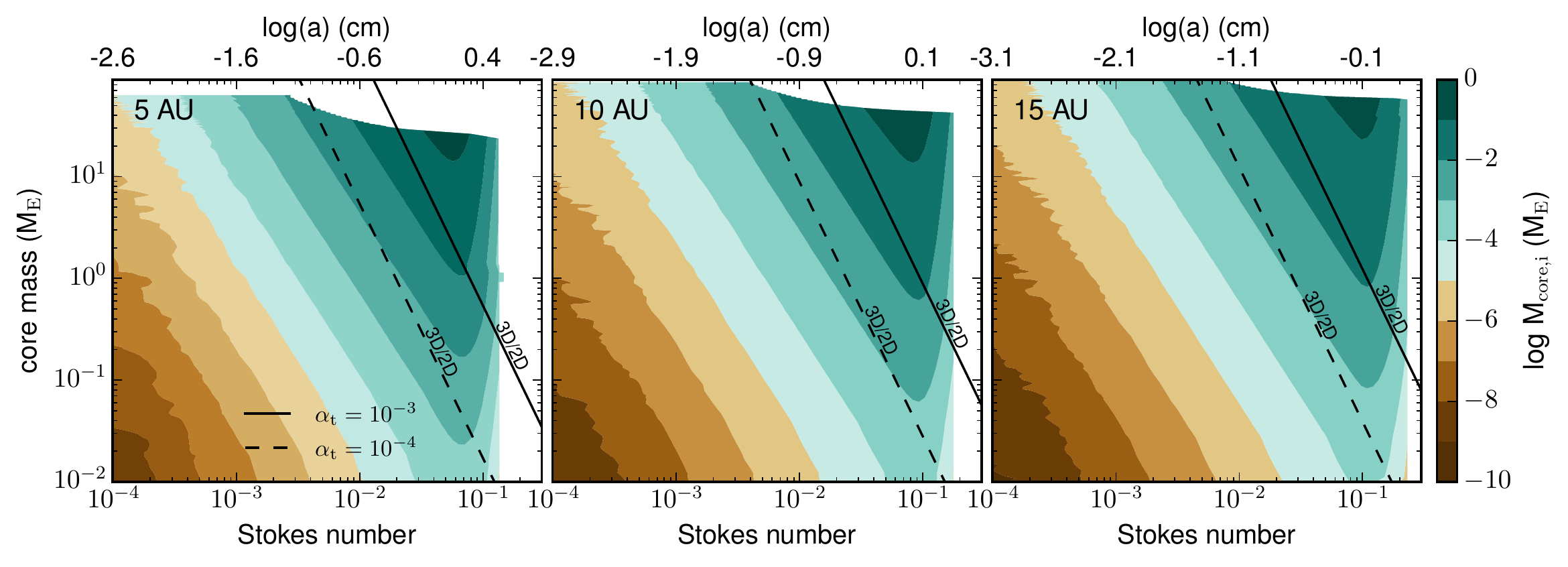} 
\caption{The planet mass at which pebble accretion switches from 3-D to 2-D as function of the Stokes number, for $\alpha_{\mathrm{t}}=10^{-3}$ and $\alpha_{\mathrm{t}}=10^{-4}$. The colour scale shows the mass $M_{\rm{core, i}}$ accumulated by the planet from the $i\rm{-th}$ pebble species, which is what each pebble species contributed to the final planet mass.
}
\label{fig:fig003}
\end{figure*}

In Figure~\ref{fig:fig003}, we illustrate the mass at which a growing planet switches from 3-D to 2-D accretion as a function of pebble size at 5 AU, 10 AU and 15 AU for turbulent strengths of $\alpha_{\mathrm{t}}=10^{-3}$ and $\alpha_{\mathrm{t}}=10^{-4}$. In the first place, the switch from 3-D to 2-D in discs with low turbulence occurs at a lower core mass, and the converse is true for high disc turbulences. This is because in less turbulent disc the pebble scale heights are low, which increases the $r_{\mathrm{H}}/h_{\mathrm{i}}$ ratio at a fixed $r_{\mathrm{H}}$ and hence constant core mass. Secondly, the accretion of smaller pebbles transitions from 3-D to 2-D mode at higher core masses. For the case of larger pebbles, this transition occurs at much lower core masses. This is because, for the same turbulence strength, larger pebbles typically settle closer to the mid-plane and hence have low scale heights compared with the smaller pebbles.

Also in Figure~\ref{fig:fig003} the contribution of particles of a given Stokes number to the total mass is shown by the colour plot, computed using Equation~(\ref{eq:CoreGrowthRegime}). Here, pebble distributions were taken from grain size reconstruction method described in more detail in Section~\ref{simulations}. From the illustration in Figure~\ref{fig:fig003}, pebbles with smaller Stokes number do not significantly contribute to the total core mass. For example, at 5 AU, pebbles with Stokes number less than 0.001 contribute roughly less than $10^{-4}~M_{\rm{E}}$ to a planetary core of 40 $M_{\rm{E}}$. This is because the smaller pebbles are not easily accreted compared with the large pebbles which carry most of the mass.

\subsection{Pebble isolation mass} \label{sec:pebbleisolation}
As the core grows massive enough, at some point it will begin to open a gap and induce a pressure bump at the outer edge of the gap where pebbles get trapped~\citep{morbidelli2012b,paardekooper2006}. Consequently, at a critical mass usually referred to as the pebble-isolation mass, the core stops accreting pebbles~\citep[\citetalias{lambrechts2014a, bitsch2018};][]{ataiee2018}. 

However, recent hydrodynamical simulations suggest that pebbles with small Stokes number can still cross to the inner disc through a gap carved out by a Jupiter-mass planet \citep[][ and references therein]{weber2018}, where pebbles with larger Stokes number are more efficiently filtered out. Since the pressure bump scales with planet mass, where typical pebbles isolation masses are an order of magnitude lower than a Jupiter-mass planet~\citep{morbidelli2012b, lambrechts2014a,bitsch2018}, the growing cores may not efficiently filter smaller pebbles. The smaller pebbles may thus overcome the pressure and hence be accreted.

The pebble isolation mass as originally developed by~\citetalias{lambrechts2014a} without consideration of turbulent effects is given by
\begin{equation}
 M_{\rm{iso}}=20\left(\frac{H/r}{0.05} \right)^{3}\rm{M_{\rm{E}}} \label{eq:lambrechts} \text{ .}
\end{equation}
\citetalias{bitsch2018} and \citet{ataiee2018} built on~\citetalias{lambrechts2014a} and investigated pebble isolation mass taking into account turbulent diffusion and obtained similar results where turbulence can significantly change pebble isolation mass. One could use either formulation to study core growth by pebble accretion, where the resulting core masses should only differ by a factor of 1.5 -- 2 as discussed in \citet{ataiee2018}. However, the \citetalias{bitsch2018} and ~\citetalias{lambrechts2014a} prescriptions  give very similar results in the limit of weak turbulence as compared with \citet{ataiee2018}. Hence for consistence we follow \citetalias{bitsch2018} since we have used classical model of~\citetalias{lambrechts2014a} to compare the final core masses with and without diffusion of pebbles across the pressure bump. \citetalias{bitsch2018} derived an expression for pebble isolation mass with diffusion as
\begin{equation}
 M_{\rm{iso}}=M_{\rm{iso}}^{\dagger}+\frac{\Pi_{\rm{crit}}}{\lambda}~M_{\rm{E}} \label{eq:bitsch1} \text{ ,}
\end{equation}
where $M_{\rm{iso}}^{\dagger}$ is the pebble isolation mass without diffusion, $\Pi_{\rm{crit}}$ is the critical pressure gradient parameter, $\lambda$ defines the change in slope of pressure gradient inside the pressure bump generated by the planet. $M_{\rm{iso}}^{\dagger}$, $\Pi_{\rm{crit}}$ and  $\lambda$  are given by
\begin{equation}
 M_{\rm{iso}}^{\dagger}=25f~M_{\rm{E}},~\lambda \approx \frac{0.00476}{f},~\Pi_{\rm{crit}}=\frac{\alpha_{\rm{t}}}{2\tau_{\rm{i}}}.\label{eq:bitsch2}
\end{equation}
Here $f$ is a fit to the isolation mass given by
\begin{equation}
f=\left[\frac{H/r}{0.05} \right]^{3}\left[0.34\left(\frac{\rm{log}(\alpha_{3})}{\rm{log}(\alpha_{\rm{t}})} \right)^{4}+0.66 \right]\left[1-\frac{\frac{\partial {\rm{ln}}P}{\partial {\rm{ln}}r}+2.5 }{6} \right]\text{,} \label{eq:bitsch3}
\end{equation}
 where $\alpha_{3}=0.001$ is the scaling factor and $\partial {\rm{ln}}P/\partial {\rm{ln}}r$ is the pressure gradient.

From Equations~(\ref{eq:bitsch1}), (\ref{eq:bitsch2}) and (\ref{eq:bitsch3}), we can calculate the pebble isolation mass with turbulent diffusion, $M_{\rm{iso,i}}$, for each pebble species as
\begin{multline}
 M_{\rm{iso,i}}=17.51\left( \frac{H/r}{0.05} \right)^{3}\left[0.34\left( \frac{\rm{log\alpha_{3}}}{\rm{log\alpha_{\rm{t}}}} \right)^{4}+0.66 \right]\\
 \times\left(3.5-\frac{\partial{\rm{ln}}P}{\partial{\rm{ln}}r} \right)\left( 0.238+\frac{\alpha_{\rm{t}}}{\tau_{\rm{i}}}\right)~M_{\rm{E}}. \label{eq:PIM}
\end{multline}

\begin{figure}
\includegraphics[width=0.5\textwidth]{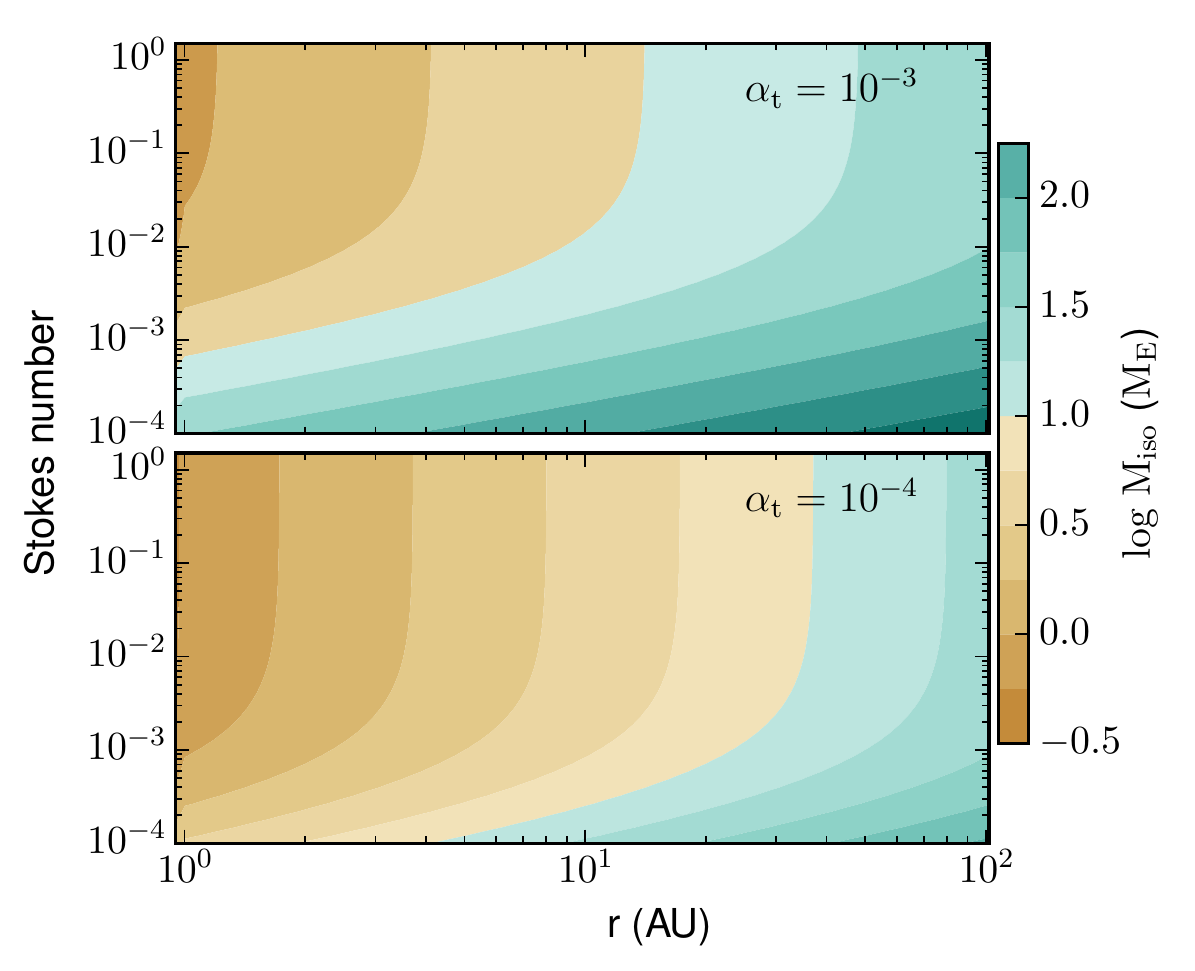} 
\caption{The pebble isolation mass as a function of pebble size and orbital distance for nominal disc turbulence parameters $\alpha_{\mathrm{t}}=10^{-3}$ and $\alpha_{\mathrm{t}}=10^{-4}$.}
\label{fig:fig004}
\end{figure}

To put Equation~(\ref{eq:PIM}) in perspective, the dependence of pebble isolation mass on grain size, turbulence viscosity and orbital distance is illustrated in Figure~\ref{fig:fig004}. For each grain species, the isolation mass increases with orbital distance, and with turbulence levels. Additionally, at a particular radial distance, pebble isolation mass increases with decreasing Stokes number since from Equation~(\ref{eq:PIM}), pebble isolation mass is inversely related to the Stokes number.

\begin{figure}
\includegraphics[width=0.5\textwidth]{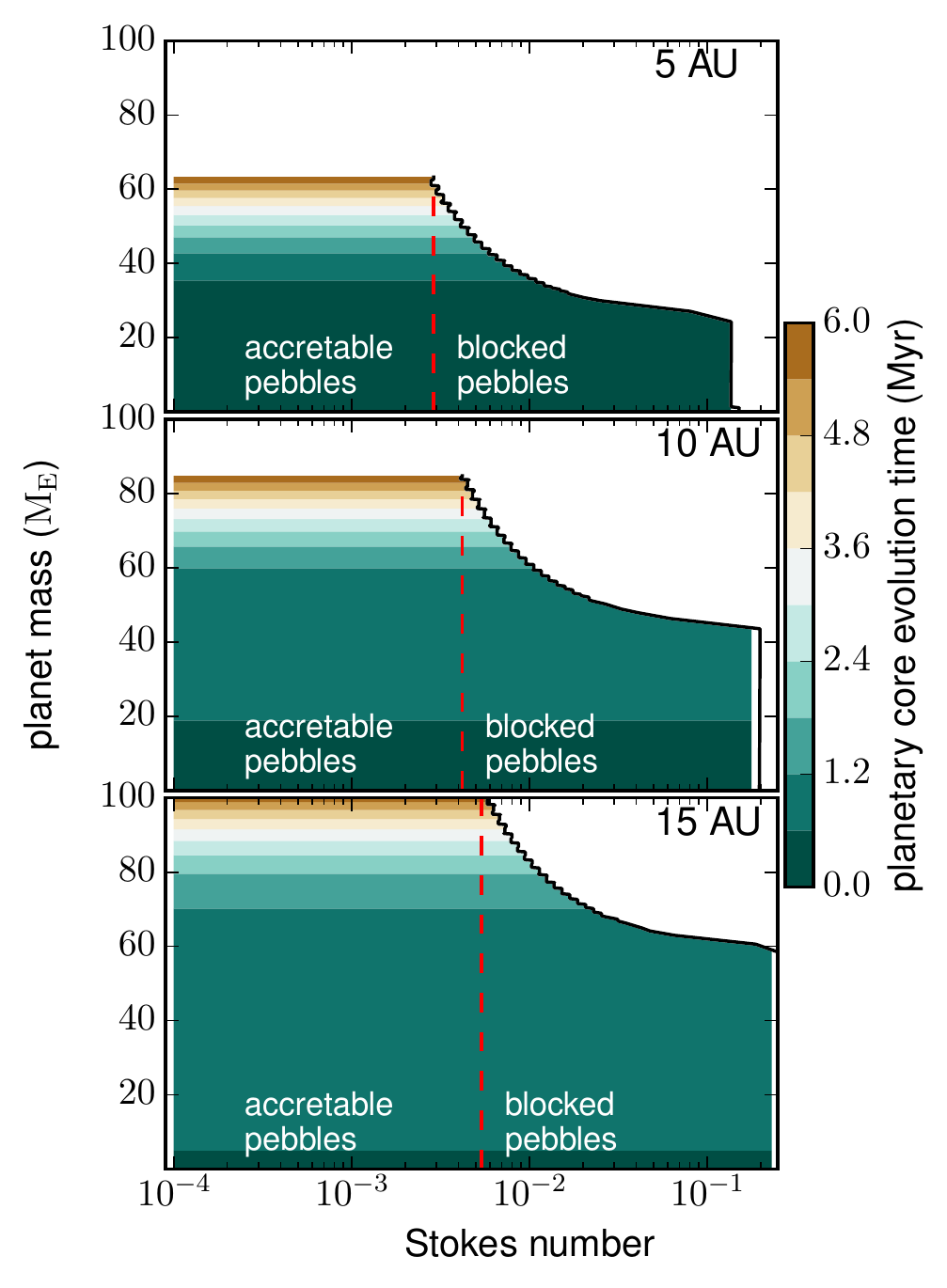} 
\caption{An illustration of how, using Equation~(\ref{eq:PIM}), a growing and non-migrating planet may filter pebbles with different Stokes numbers at 5 AU, 10 AU and 15 AU for a nominal disc turbulence parameter of $\alpha_{\mathrm{t}}=10^{-3}$, fragmentation velocity of 10 m/s  and dust-to-gas ratio of 0.01, where grain sizes were obtained from the method described in Section~\ref{simulations}.}
\label{fig:fig002}
\end{figure}

As a consequence of Equation~(\ref{eq:PIM}), a growing planet may block pebbles at different stages as illustrated in Figure~\ref{fig:fig002}, which shows what pebble species a non-migrating planet accretes as it reaches different masses. For example, at 5 AU a 20 $M_{\mathrm{E}}$ planet is accreting all pebble species. When the planet reaches 60 $M_{\mathrm{E}}$, it has blocked all pebble species to the right of the dashed vertical line and can only accrete pebbles to the left of the vertical line. At higher orbital distances, the planet needs to grow bigger before it can start filtering larger pebbles. 

\subsection{Planetary migration scheme}
We implement orbital evolution during core growth and follow the~\citet{paardekooper2011} prescription, where cores experience Lindblad and corotation torques $\Gamma_{\rm{L}}$ and $\Gamma_{\rm{C}}$, respectively with the total torque $\Gamma_{\rm{tot}}$ given by
\begin{equation}
 \Gamma_{\rm{tot}}=\Gamma_{\rm{L}}+\Gamma_{\rm{C}}.\label{torque1}
\end{equation}
Here, the Lindblad torque is expressed as
\begin{equation}
 \frac{\gamma\Gamma_{\rm{L}}}{\Gamma_{0}}=-2.5-1.7\beta+0.1s,
\end{equation}
where $\gamma=1.4$ is the adiabatic index, $s$ and $\beta$  are the negatives of radial gradients of gas surface density, $\Sigma_{\rm{pla}}$, and the temperature, $T$, respectively calculated at the planet's location, $r_{\rm{pla}}$.  $\Gamma_{0}$ is expressed as
\begin{equation}
 \Gamma_{0}=\left(\frac{q}{H/r}\right)^{2}\Sigma_{\rm{pla}}r_{\rm{pla}}^{4}\Omega_{\rm{pla}}^{2},
\end{equation}
where $q$ is the planet-star mass ratio and $\Omega_{\rm{pla}}$ the Keplerian frequency and  the disc aspect ratio $H/r$ is calculated at planet's location.
The corotation torque which is induced by material corotating with the planetary body is composed of barotropic and entropy related parts. For detailed discussion, we refer the reader to the review by~\citet{baruteau2014}. The corotation torque is calculated using the formula~\citep{paardekooper2010}
\begin{equation}
 \frac{\gamma\Gamma_{\rm{C}}}{\Gamma_{0}}=1.1\left(\frac{3}{2}-s \right)+ 7.9\frac{\xi}{\gamma},
\end{equation}
where the first and second terms are the barotropic and entropy components of the corotation torque and $\xi=\beta-(\gamma - 1)s$ is the radial entropy gradient.

In our model, we only incorporated type~I migration but not type~II because our model is limited to solid core growth just before pebble isolation mass, making type~II migration not necessary.
\section{Numerical simulations}\label{simulations}
The full dust size population in~\cite{birnstiel2010, birnstiel2011, birnstiel2012} features a large number of different dust species,  broadly classified into small and large dust size population. Some of these species are either accretable or just contribute to the disc opacity, which then determines the temperature profile of the disc.

We incorporated the two-population dust evolution code of~\cite{birnstiel2012}\footnote{\url{https://github.com/birnstiel/two-pop-py}} in our numerical code for dust evolution. In the simulations, particles evolve both in time and space as the disc evolves, governed by balance between grain growth, fragmentation and drift size limits. We then reconstructed the surface density of each particle species for a sample of 150 species at every time step, using the size distribution reconstruction code of~\cite{birnstiel2015}\footnote{\url{https://github.com/birnstiel/Birnstiel2015_scripts}}. 

We performed our simulations in axisymmetric 1-D disc, with 200 logarithmically spaced radial grid points. For the global disc evolution, the computational grid extends from 0.05 -- 3000 AU with characteristic radius, $r_{\rm{c}}=200$ AU, while we implant 0.01 $M_{\rm{E}}$ planetary embryos between 1 -- 50 AU. The central star has mass, $M_{*}$ = 1.0 $M_{\odot}$, temperature, $T_{*}$ = 5778 K, radius, $R_{*}$ = 1.0 $R_{\odot}$. We assume a disc mass, $M_{\rm{disc}}$ = 0.1 $M_{\odot}$, which gives $\sim 330~M_{\rm{E}}$ of dust mass for a nominal solid-to-gas ratio of 0.01. This is within the range of dust masses measured in different star-forming regions, especially for some Class 0 disc systems~\citep[see, e.g.,][]{manara2019, tychoniec2020}.

In the the simulations, we tested three different initial dust-to-gas ratios, $f_{\rm{DG}}$ = 0.01,0.015, 0.02 with the following combinations of fragmentation velocity and turbulence parameter: 
\begin{enumerate}[--]
 \item $u_{\rm{f}}=10~\rm{m/s},~\alpha_{t}=10^{-3}$;
 \item $u_{\rm{f}}=10~\rm{m/s},~\alpha_{t}=10^{-4}$;
 \item $u_{\rm{f}}=1~\rm{m/s},~\alpha_{t}=10^{-3}$;
 \item $u_{\rm{f}}=1~\rm{m/s},~\alpha_{t}=10^{-4}$.
\end{enumerate}
Here, the turbulence parameter $\alpha_{\mathrm{t}}$ regulates grain size in the fragmentation regime, pebble scale height and migration, which all change at the same time when $\alpha_{\mathrm{t}}$ changes. The disc temperature is assumed to be constant in time and varies only with radial distance as defined in the two-population code.

We performed two sets of simulations. In the first part, we adopted the classical core accretion of dominant pebble size where the isolation mass is governed by the classical Equation~(\ref{eq:lambrechts}) of~\citetalias{lambrechts2014a}. In our dominant species model, we used the full size distribution to calculate the mass averaged Stokes number and the surface density corresponding to the dominant size, as described in Section~\ref{sec:dominantspecies}. 
 
 In the second part, we feed the full size distribution into the core accretion routine, where pebbles of a given Stokes number are accreted independently. In this scenario, pebble isolation mass of the individual species is set by Equation~(\ref{eq:PIM}) of~\citetalias{bitsch2018}, and we calculate the accretion rates according to the recipe described in Appendix~\ref{sec:appendixB}. Furthermore, we terminate the accretion of full size distribution and measure the planet mass when the core has reached the isolation mass of pebble species with Stokes number $\approx$ 0.001. This is because, as we saw in Figure~\ref{fig:fig003}, the contribution of pebbles with Stokes number less than 0.001 is small and stopping their accretion after the isolation mass of pebbles with Stokes number $\approx$ 0.001 is reached will not significantly affect our results.

 Migration of planetary cores also take place during core growth. We terminate core growth when the planet has migrated to 0.05 AU or when it has reached the pebble isolation mass. If the core fails to reach pebble isolation mass, growth continues until the end of time evolution of the disc.

\section{Results and discussion}\label{results}
\begin{figure*}
\includegraphics[width=\textwidth]{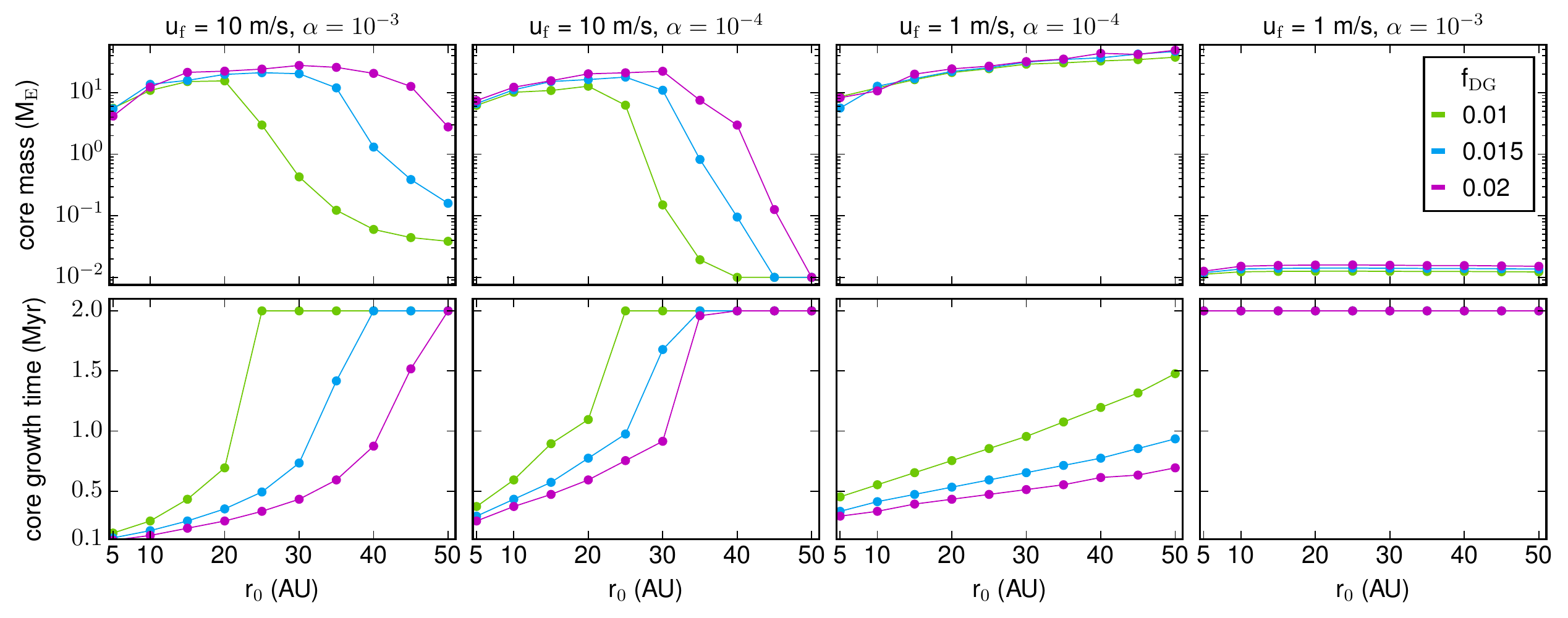} 
\caption{\textit{Top panel:} The final core masses as a function of the starting position resulting from accretion of dominant pebble species. The dominant pebble size is calculated from the reconstructed pebble size distribution as derived from the full dust evolution model prescribed in~\citet{birnstiel2012}. Core growth is measured within a period of 2 Myr of gaseous disc life span for different combinations of fragmentation velocity and turbulence levels. The final planetary cores become increasingly more massive if the dust-to-gas ratio $f_{\rm{DG}}$ increases from 0.01 to 0.02. \textit{Bottom panel:} The corresponding total growth times of the planetary cores for the same values of parameters on the left panel. The planetary cores with growth times below 2 Myrs reached their pebble isolation mass, and the growth time to reach isolation mass reduces with increasing dust-to-gas ratios.}
\label{fig:fig01}
\end{figure*}

We present and discuss results for the two pebble accretion models. In the first part, we start by discussing the classical scenario of core growth with dominant size distribution, where the isolation mass is determined by Equation~(\ref{eq:lambrechts}). Here, the dominant sizes were obtained by computing the mass averaged dust size and Stokes number from the full grain size distribution reconstructed from the two-population model~\citep{birnstiel2012}. In the second part we present results of our model featuring concurrent accretion of multi-pebble species, where pebble isolation mass is controlled by Equation~(\ref{eq:PIM}). We then point out the differences between the two models and discuss what it means for formation of planetary systems by core accretion paradigm in relation to current observations. Although the transition mass is sensitive to initial position of the embryo, throughout our work, we assumed embryos with transition mass $0.01 M_{\rm{E}}$ for all initial orbital positions considered here.

\subsection{Core growth with dominant pebble species}\label{sec:dominant}
 
Figure~\ref{fig:fig01} represents core growth timescales and final masses that were obtained from core growth by accretion of dominant size. Here, we show results for different sets of fragmentation velocity and turbulence parameter. The plots in top and bottom rows respectively represent the final core masses and total growth times as a function of starting positions of the embryos. Here, we first determined at what initial orbital positions our planetary embryos can grow and reach pebble isolation mass. From the plots, the planetary cores that took 2 Myrs had not reached their pebble isolation mass. We note here that migration of planetary cores also took place during the entire process of core growth (their final orbital distances and growth times will be discussed in more detail later on in Section~\ref{sec:growthtacks}).

From Figure~\ref{fig:fig01}, our simulations with $u_{\rm{f}}=1~\rm{m/s}$ and $\alpha_{\rm{t}}=10^{-3}$ produced no significant growth of planetary embryos over the 2 Myr of disc evolution. This is because low fragmentation velocities in turbulent discs keep overall grain sizes small and more coupled to the gas, which makes their accretion difficult. Accretion becomes even more difficult when planetary cores are very small, where small dust grains simply drift past the embryo~\citep{guillot2014}. This is because the small embryos do not have strong enough gravitational force to pull off small grains that are strongly attached to the gas. 

For the simulations performed with fragmentation velocity of $10~\rm{m/s}$, and turbulence strengths of $10^{-3}$ and $10^{-3}$, planetary cores reached pebble isolation mass only when growth of the embryos started at a radial distance within 20 AU for nominal dust-to-gas ratio of 0.01. The corresponding growth timescales for the cores implanted within 20 AU to reach their isolation mass ranges between 0.1 -- 1.5 Myrs as shown in the bottom panels of Figure~\ref{fig:fig01}. 

With increasing dust-to-gas ratio, core growth is boosted and planetary embryos that are introduced as far as 35 AU can now reach the pebble isolation mass. At each initial radial distance, the growth timescales are also greatly improved when the dust-to-gas ratio increases. The improved growth rates at higher dust-to-gas ratios  can be related to the higher dust surface densities that allow cores to grow faster and bigger. Furthermore, radial drift motions of dust also depend on dust-to-gas ratio, which when increased, can bring about fast inward transport of solid material from the outskirts of the disc. This then increases pebble flux and  accretion efficiency at the planet's location, and hence the embryos may effectively grow bigger with reduced growth time.

From Figure~\ref{fig:fig01} where $\alpha_{\rm{t}}=10^{-4}$ and $u_{\rm{f}}=1~\rm{m/s}$, we registered much better final core masses in comparison with the other models. Here, planetary cores can grow to 20 -- 30 $M_{\rm{E}}$ when the cores start growing at orbital distances beyond 20 AU.  This is contrary to what has so far been reported in most pebble accretion models based on the dominant size approach, where growing such massive cores when planetary embryos start at wider orbits is difficult. Furthermore, many of these studies have predicted that pebble accretion is either too efficient or inefficient, depending on the prevailing physical conditions in the disc. However, in our simulations, we may attribute such massive core sizes to the fact that low fragmentation velocities keep grain sizes small. These small size grains migrate to the inner disc regions slowly compared with larger grains and may last longer in many parts of the disc, thereby promoting core growth especially large orbital distances.

In our simulations, the growth timescales to reach pebble isolation mass typically span from $\sim$ 0.2 to 1.8 Myr as shown in the bottom panel of Figure~\ref{fig:fig01}. This depends on the initial location of embryo and the dust-to-gas ratio, where in the inner disc regions the cores take shorter time to grow to their isolation mass. This is because the pebble isolation mass as found by~\citetalias{lambrechts2014a} and~\citetalias{bitsch2018} is a cubic function of the disc aspect ratio, which increases with orbital distances. Thus, cores that grow at far orbital distances require more time and material to reach isolation mass~\citep[see][for a more detailed discussion]{bitsch2015b}.

\subsection{Core growth with full grain size distribution}\label{sec:fullsize}
\begin{figure*}
\includegraphics[width=\textwidth]{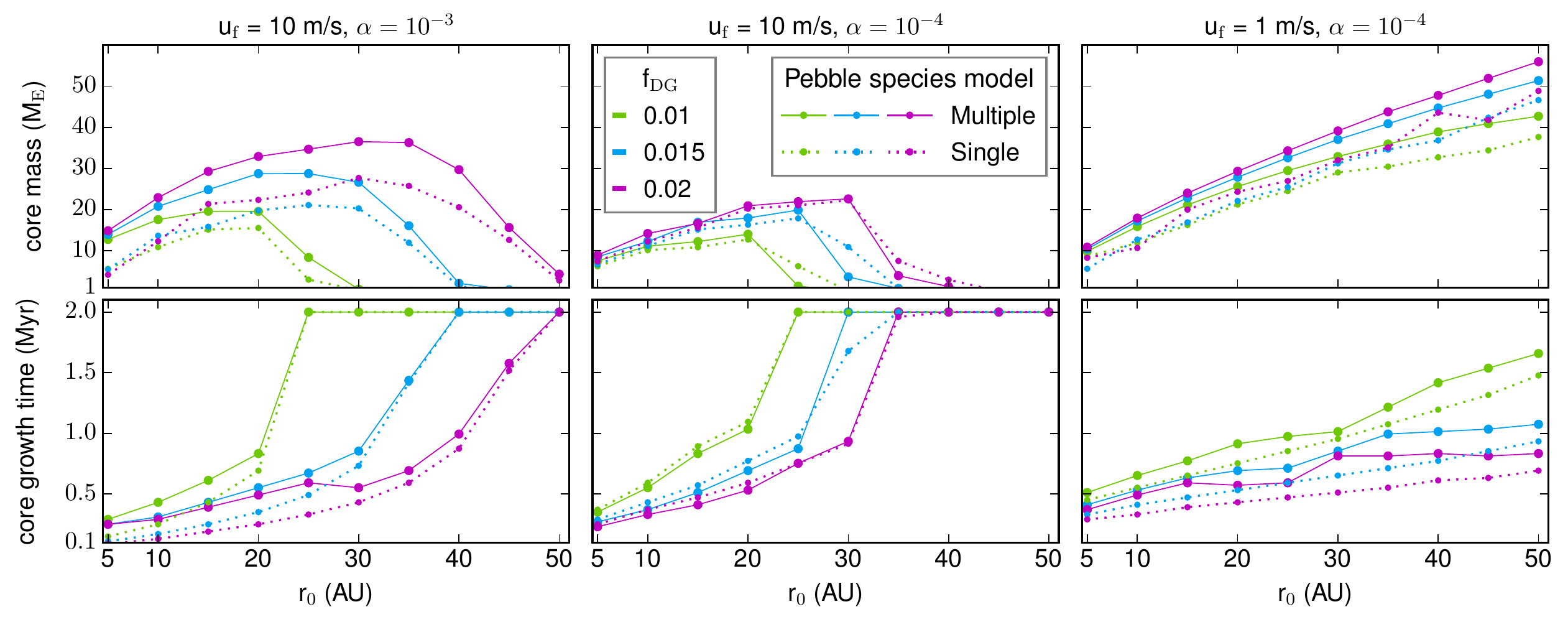} 
\caption{The final core masses (\textit{top panel}) and the total growth time (\textit{bottom panel}) as a function of the starting position, resulting from concurrent accretion of multiple pebble species considering full pebble size distribution. The plots have the same meaning as in Figure~\ref{fig:fig01}. Here the same single species model from Figure~\ref{fig:fig01} is overplotted for comparison with concurrent accretion of full size distribution model. The planets in these plots migrate and their growth tracks are shown in Figure~\ref{fig:fig05}.}
\label{fig:fig03}
\end{figure*}

\begin{figure*}
\includegraphics[width=\textwidth]{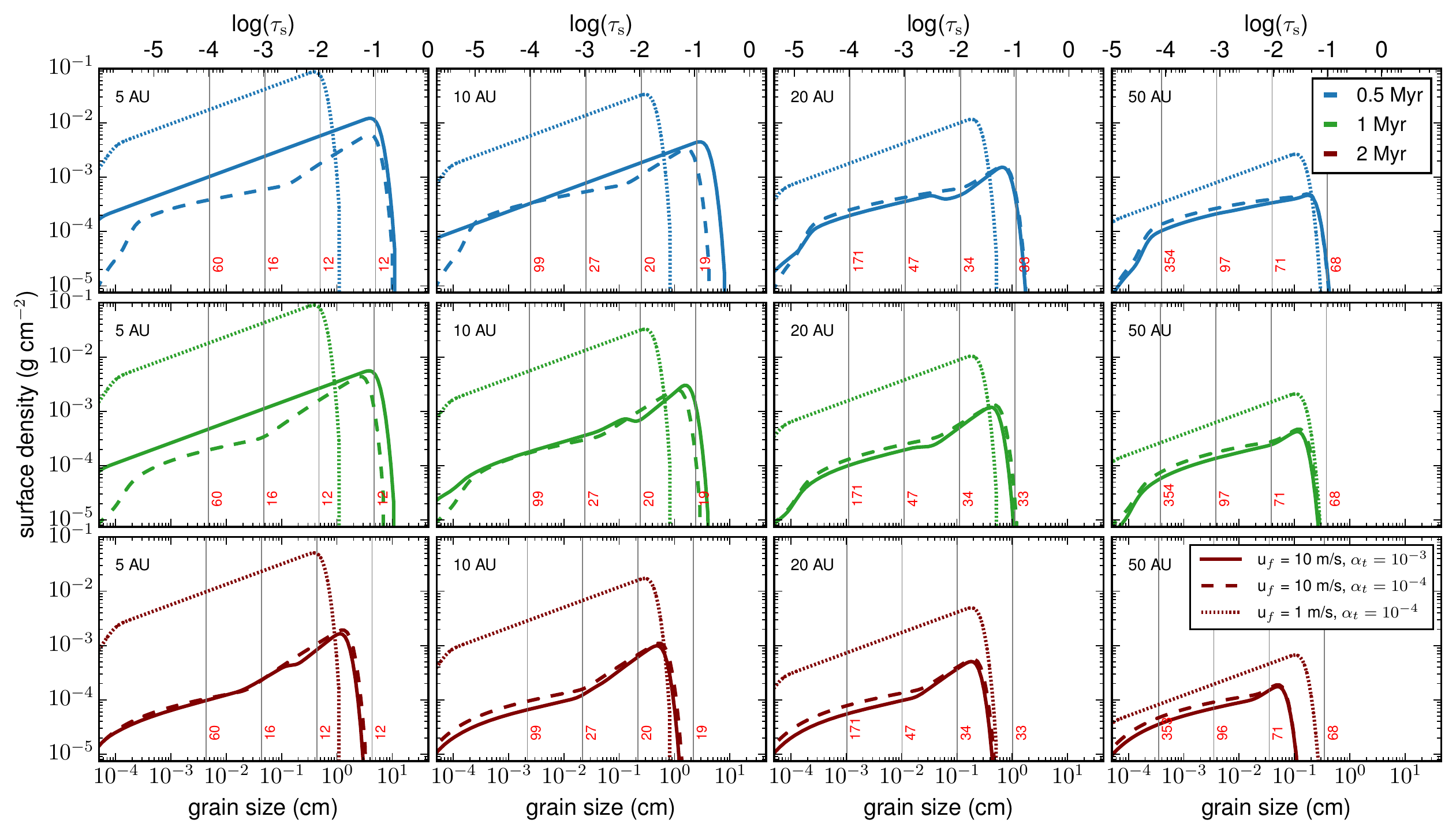} 
\caption{Grain size distribution reconstructed after  0.5 Myr (top row), 1 Myr (middle row) and 2 Myr (bottom row) of disc evolution for nominal $f_{\rm{DG}}=0.01$. We derived the grain size distributions using $\alpha_{\rm{t}}=10^{-3}$ with $u_{\rm{f}}$ = 10 m/s (solid line), $\alpha_{\rm{t}}=10^{-4}$ with $u_{\rm{f}}$ = 10 m/s (dashed line) and $\alpha_{\rm{t}}=10^{-4}$ with $u_{\rm{f}}$ = 1 m/s (dotted line). The Stokes numbers corresponding to the grain sizes are indicated on the upper axis. The grain sizes were reconstructed from the two-population model of~\citet{birnstiel2012} where dust evolution is governed by growth, fragmentation and drift limits. Here, the large population carries 0.75 and 0.97 of the solid mass in the fragmentation and drift limits, respectively (for details, see Appendix~\ref{sec:appendixA}). The vertical lines show pebble species that would be blocked at a planetary mass labelled on each line. The line labels are pebble isolation masses with diffusion corresponding to $\alpha_{\mathrm{t}}=10^{-4}$. Here, all pebble species on the right of the vertical line would be blocked.}
\label{fig:fig02}
\end{figure*}

\begin{figure*}
\includegraphics[width=\textwidth]{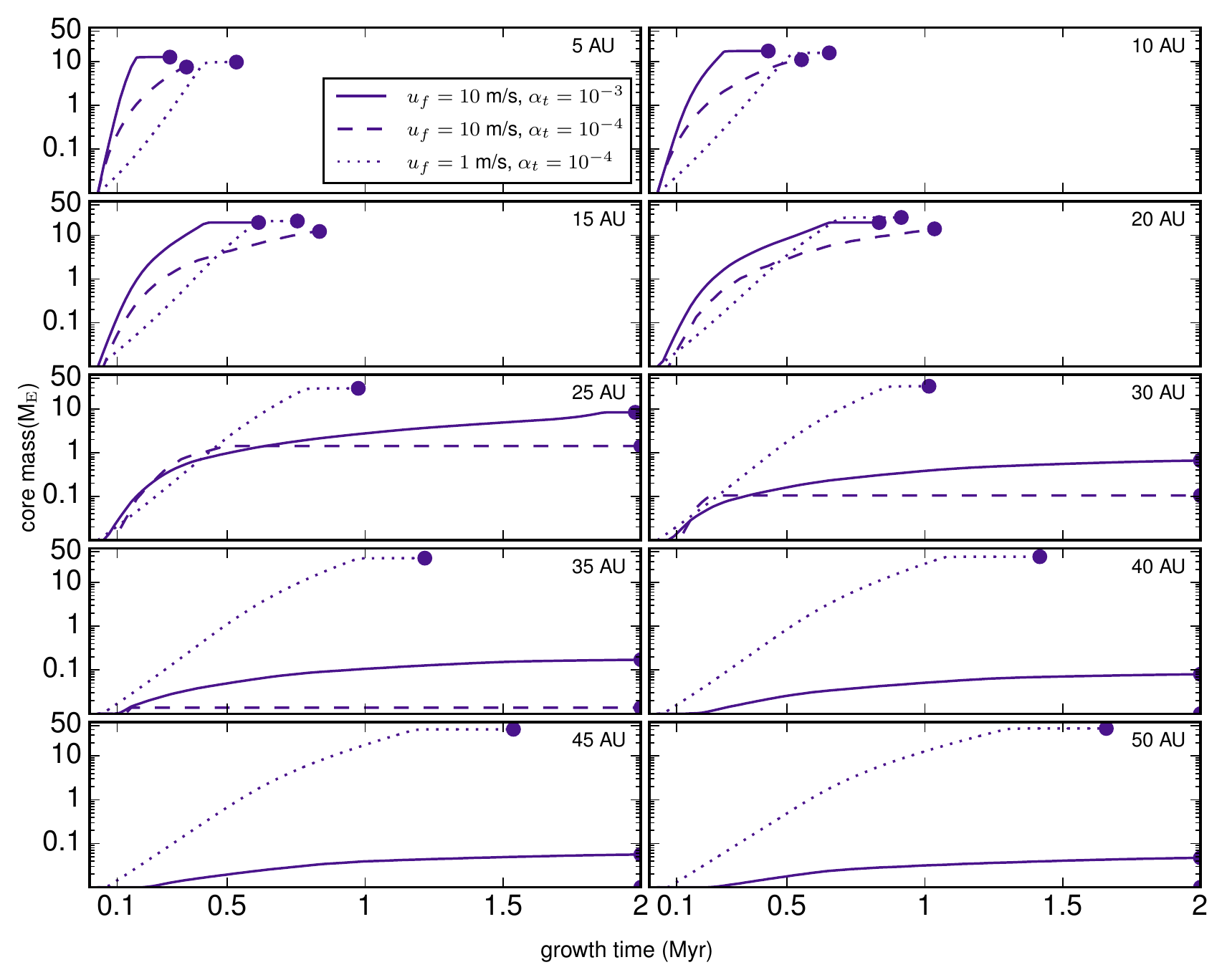} 
\caption{The growth time of the cores in Figure~\ref{fig:fig03} that grow by accretion of multi-pebble species paradigm for nominal $f_{\rm{DG}}=0.01$, with the starting positions of planetary embryos shown on each plot. Some planetary cores grow very fast initially, and  then the growth quickly stalls as shown by the horizontal lines. Along the horizontal line, the planet is accreting pebble species with the smallest Stokes numbers which have insignificant contribution to the core mass.}
\label{fig:fig04}
\end{figure*}

In Figure~\ref{fig:fig03}, we present the planetary core masses resulting from accretion of full dust size distribution, where the size of different dust species were reconstructed using the grain size reconstruction recipe of ~\citet{birnstiel2015}. In the simulations, we used the pebble isolation mass with diffusion to separate pebbles species for which the isolation mass has been attained. This ensures that when the planet has reached isolation mass of a particular pebble species, that species is not accreted again during the rest of core growth period. In Figure~\ref{fig:fig03}, we overplotted the results of the dominant species model presented in Figure~\ref{fig:fig01} for purpose of comparison with the multi-species model.

From Figure~\ref{fig:fig03}, in the multi-pebble species approach, the core masses increase substantially in comparison with our dominant species approach. This is because the cores accrete not just a single but a variety of pebble species, where growth is sustained by accretion of smaller pebbles even after the larger pebbles have been isolated. We note here that the sustained growth is a consequence of each pebble species being isolated at different core masses due to the dependence of pebble isolation on pebbles Stokes number (see Equation~(\ref{eq:PIM})). Consequently, the core takes more time to grow than in the case of single species model. In the latter case, accretion stops as soon as the planet just reaches its classical pebble isolation mass, and hence the planets have shorter growth times.  We shall discuss this in detail in Section~\ref{sec:PIM}.

Keeping the fragmentation velocity at $10~\rm{m/s}$ and reducing the turbulence strength to $\alpha_{\rm{t}}=10^{-4}$, results into final core masses that are much smaller than those obtained from the model with turbulence level $\alpha_{\rm{t}}=10^{-3}$. Here, the growth patterns in both the single and multi-pebble species models are very similar.
This is because the low turbulence level has three main effects: first, it allows larger grain size distribution which settle more efficiently toward the mid-plane; secondly, the larger grains drift much faster than the smaller grains produced in a more turbulent disc. The former effect would result in better accretion efficiency, but then this could be counteracted by fast drifting grains which could result in significant loss of the larger grain. Thus, all in all, for $10~\rm{m/s}$ and $\alpha_{\rm{t}}=10^{-4}$, the cores may accumulate much smaller material and take a lot more time to reach pebble isolation mass in such an environment. Thirdly, low turbulence levels reduce turbulent diffusion of grains across the pressure bump. Hence in this case, the pebble isolation mass with diffusion approaches the classical case without diffusion~\citep{bitsch2018, ataiee2018}. Here, the pebble isolation mass is smaller than the case where turbulence strength is greater, hence resulting into smaller core masses.  This is further illustrated in Figure~\ref{fig:fig02}, where for $\alpha_{t}=10^{-4}$, pebbles with Stokes number greater 0.01 can be isolated by very similar planet masses (shown by red labels on the vertical lines). This also explains why the final core masses in both the dominant and multi-species models in the middle panel of Figure~\ref{fig:fig03} are very similar.

With $u_{\mathrm{f}}=1~\rm{m/s}$ and $\alpha_{t}=10^{-4}$, we obtained supermassive cores in the same way as the case of single species model as shown on the right panel of Figure~\ref{fig:fig03}. To further understand this trend, we refer to the grain size distributions shown in Figure~\ref{fig:fig02}.
Firstly, as demonstrated in Figure~\ref{fig:fig02}, a fragmentation velocity of 1 m/s keeps pebbles at rather much smaller sizes, and so drift at lower speeds than larger pebbles produced when the fragmentation limit is set to 10 m/s. Consequently, the small and slow drifting pebbles live much longer (possibly everywhere in the disc) than the larger pebbles which may drain much more quickly. This creates opportunity for core growth at wider orbits. For example, from Figure~\ref{fig:fig02}, the grain size distribution remains more stable after 1 Myr of disc evolution for our disc model with $\alpha_{\rm{t}}=10^{-4}$ and $u_{\rm{f}}=1~\rm{m/s}$ compared with the other two models.  Secondly, the low turbulence ensures that the small pebbles are not stirred too far from the mid-plane.

From the above discussion, it implies that in  the disc model with $\alpha_{\rm{t}}=10^{-4}$ and $u_{\rm{f}}=1~\rm{m/s}$, pebbles are kept within the feeding zone of the core, possibly most of the time. The cores can then continue to accrete the small pebbles for an extended period of time. This suggests that at low disc turbulence and low fragmentation velocities, core growth from pebble accretion might be possible even in the last stages of the disc lifetime.

We note that the classical dominant species approach relies on the premise that the grain population containing the biggest solid mass budget constitutes grains of very similar aerodynamic size. This can be true if turbulence, fragmentation and grain growth conspire rightly to keep grains at very similar sizes. Consequently, in such an environment where grain size distribution can be similar, our multi-pebble species approach should yield  very similar core masses as the classical dominant species model, possibly at faster growth rates.

In addition, our multi-species model requires two conditions: (i) the grains need to have varied size distribution with dissimilar aerodynamic properties and (ii) turbulence is required to operate in such a way as to enforce pebble isolation mass dependence on pebble Stokes number. The first condition leads to a departure from the classical dominant species model because the different grain species are now subjected to different gas drag, which underpins pebble accretion. The second condition is required for sustaining core growth from smaller pebbles after the larger ones have been isolated. This holds only if pebble isolation mass is not supposedly universal for all pebble species at a given radial location.

In our simulations, core growth rates and the final core masses are sensitive to dust-to-gas ratio, fragmentation velocity and turbulence strength with the incorporation of the full grain size distributions. This gives a diverse outcome of final core masses as shown in Figure~\ref{fig:fig03}.

It should be noted that the masses in Figure~\ref{fig:fig02} do not directly correspond to the masses in Figure~\ref{fig:fig03}. This is because in Figure~\ref{fig:fig03}, we show the planet mass as a function of the starting position. On the other hand, the orbital distances in Figure~\ref{fig:fig02} are sample distances at which we constructed grain size distributions, and at which we showed sample isolation masses to illustrate at what mass pebble species may be blocked.  Therefore, since we incorporated orbital migration, we need the final orbital distances from Figure~\ref{fig:fig05} if we wish to relate the core masses in Figure~\ref{fig:fig03} to the isolation masses in Figure~\ref{fig:fig02}. Furthermore, according to our  multi-species accretion model, accretion stops once  pebbles with $\tau_{\rm{i}}=0.001$ are blocked as earlier explained in Section~\ref{simulations}. This means that the final core mass is determined by the pebbles with Stokes number closest to 0.001. Because this Stokes number may be slightly greater than 0.001, the final core mass may not necessarily correspond to the isolation mass for the Stokes number 0.001 indicated in Figure~\ref{fig:fig02}, but closer to it.

\subsection{Growth tracks of planetary cores}\label{sec:growthtacks}
\begin{figure}
\includegraphics[width=0.475\textwidth]{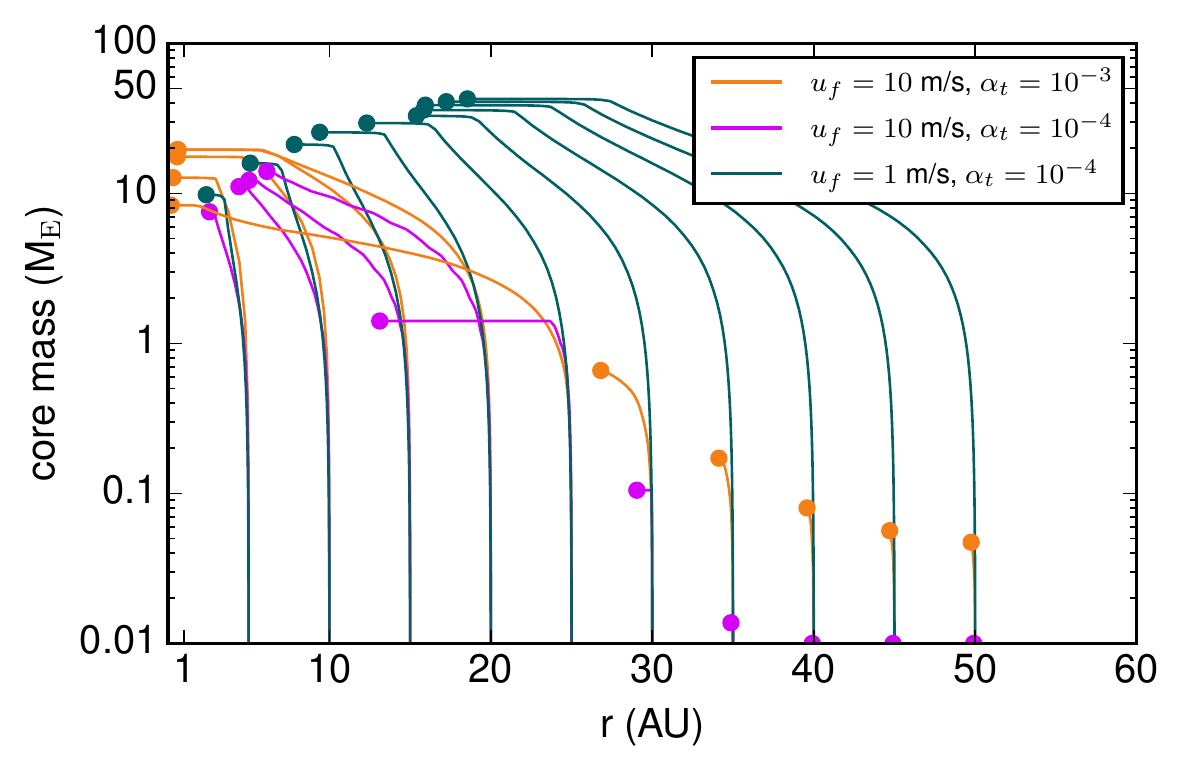} 
\caption{The growth tracks of the cores in Figure~\ref{fig:fig03} for nominal $f_{\rm{DG}}=0.01$. The solid lines show the evolution of the orbital position and the filled circles indicate the final masses and positions of the cores. The horizontal migration tracks show that the planet is accreting pebble species with smallest Stokes numbers and does not significantly increase in mass, as it continues to migrate. }
\label{fig:fig05}
\end{figure}
Figure~\ref{fig:fig04} represents the time evolution of the planetary cores  that grow via the multi-species accretion paradigm and initially implanted at orbital positions shown in Figure~\ref{fig:fig03}. The corresponding orbital evolution is shown in Figure~\ref{fig:fig05} for a nominal dust-to-gas ratio of 0.01.

In Figures~\ref{fig:fig04} and~\ref{fig:fig05}, the points where the growth tracks flatten show the points of growth saturation, and consequently there is no significant increase in mass. This occurs when the majority of pebbles species have been isolated, but the core continues to slowly accrete the remaining pebble species with the smallest Stokes number. It is important to remember that in our simulations, the final mass of the planet is determined when it reaches the isolation mass of pebbles with Stokes number $\approx$ 0.001. This then sets the final core growth time of the planets. In our results the constribution of pebbles with small Stokes number close to 0.001 is negligibly small as reported in several studies~\citep[e.g.,][]{guillot2014, johansen2019}. This is shown by the flattening of mass-time curves for the growing cores (see Figure~\ref{fig:fig04}). The growth tracks which end at 2 Myr of disc evolution pertain to cores that have not yet reached their pebble isolation masses.  

In Figure~\ref{fig:fig04}, planetary cores accrete more efficiently in the moderately turbulent disc where we set $u_{\rm{f}}=10~\rm{m/s}$ and $\alpha_{\rm{t}}=10^{-3}$ compared with the models for which the turbulence parameter is set to $\alpha_{\rm{t}}=10^{-4}$. The apparent reasons for the differences in the growth times for these models  are associated with the particle size distribution regulated by growth, fragmentation and turbulence as previously discussed in Subsection~\ref{sec:fullsize}. 

With $u_{\rm{f}}=10~\rm{m/s}$ and $\alpha_{\rm{t}}=10^{-4}$, the grain size distribution mostly constitutes larger grains {{that drift faster compared with the other two models}}. This means smaller material availability and hence slower core growth rates as most of the larger grains are lost on short timescales via radial drift. This results into a relatively slow growth and hence longer growth times as compared with the other two models as shown in Figure~\ref{fig:fig04}, especially for initial orbital positions between 10 -- 20 AU. 

The size distribution produced by  $u_{\rm{f}}=1~\rm{m/s}$ and $\alpha_{\rm{t}}=10^{-4}$ are typically in millimetre range. These smaller sized dust grains are accreted less efficiently, resulting into extended growth times before the cores reach pebble isolation mass. However, concurrent accretion of these smaller multi-pebble species allows the planet to reach pebble isolation mass well before the 2 Myr of disc evolution, even for wider initial orbital locations.

Since the final core masses are determined by the pebble isolation mass for the smaller pebbles, one would expect longer growth times because of the slow accretion rates of these small pebbles. However, initially the accretion of larger pebbles is more efficient than the smaller pebbles whose accretion rate increases as the core grows bigger. Also the concurrent accretion of different species helps the core to rapidly increase in mass in a much shorter time. Thus the smaller pebbles can now be accreted much more efficiently, thereby shortening the time for the planet to reach the isolation of the smallest species.

In Figure~\ref{fig:fig05}, we show the core growth trajectories for the same sets of the parameters as in Figures~\ref{fig:fig04}. Here, planetary cores that were able to reach the pebble isolation mass migrated significantly inward. However, the fast growth rates of some cores propel them to reach pebble isolation mass after migrating over relatively short orbital distances. For example, for the models with low turbulence level of $\alpha_{\rm{t}}=10^{-4}$, planetary cores that start at 5 AU migrated relatively short distances (by $\approx$ 2.5 AU) by the time they reached their isolation mass. Thus there may be a possibility that planetary core can form almost in-situ in some parts of the disc if core growth proceeds by concurrent accretion of multiple pebble species.

Analogous to the core growth saturation points discussed above, the points where the curves in Figure~\ref{fig:fig05} flatten mark the orbital distance at which the cores reached their isolation mass. We remind the reader here that we did not include type~II migration which might significantly change the final orbits of the planets.

In our simulations, we were interested in core growth and thus did not model gas accretion as well as transition from type~I to type~II migration. Type~II migration could be triggered during gas accretion phase after the cores have reached their full pebble isolation mass. We envision that with the inclusion of type~II migration, orbital decay is enhanced and could be significant to the extent that planets are lost to the host star, unless slower migration mechanisms like the dynamical torque~\citep{paardekooper2014} and slower type~II migrations~\citep[as in][]{crida2017,crida2017b,robert2018, bergezcasalou2020, ndugu2021} are considered. Therefore,  even though our simple multiple-species accretion model shortens the growth time of cores, the orbital dilemma of gas giant planets migration needs to be explored further in the accretion of multi-pebble species scenario. We shall investigate this in our upcoming work. However, a promising solution is the heating torque~\citep{llambay2015} which could prevent loss of planets to the central star by stopping inward migration, and this allows planet formation closer to the star.

\subsection{The role of pebble isolation mass}\label{sec:PIM}
\begin{figure}
\includegraphics[width=0.475\textwidth]{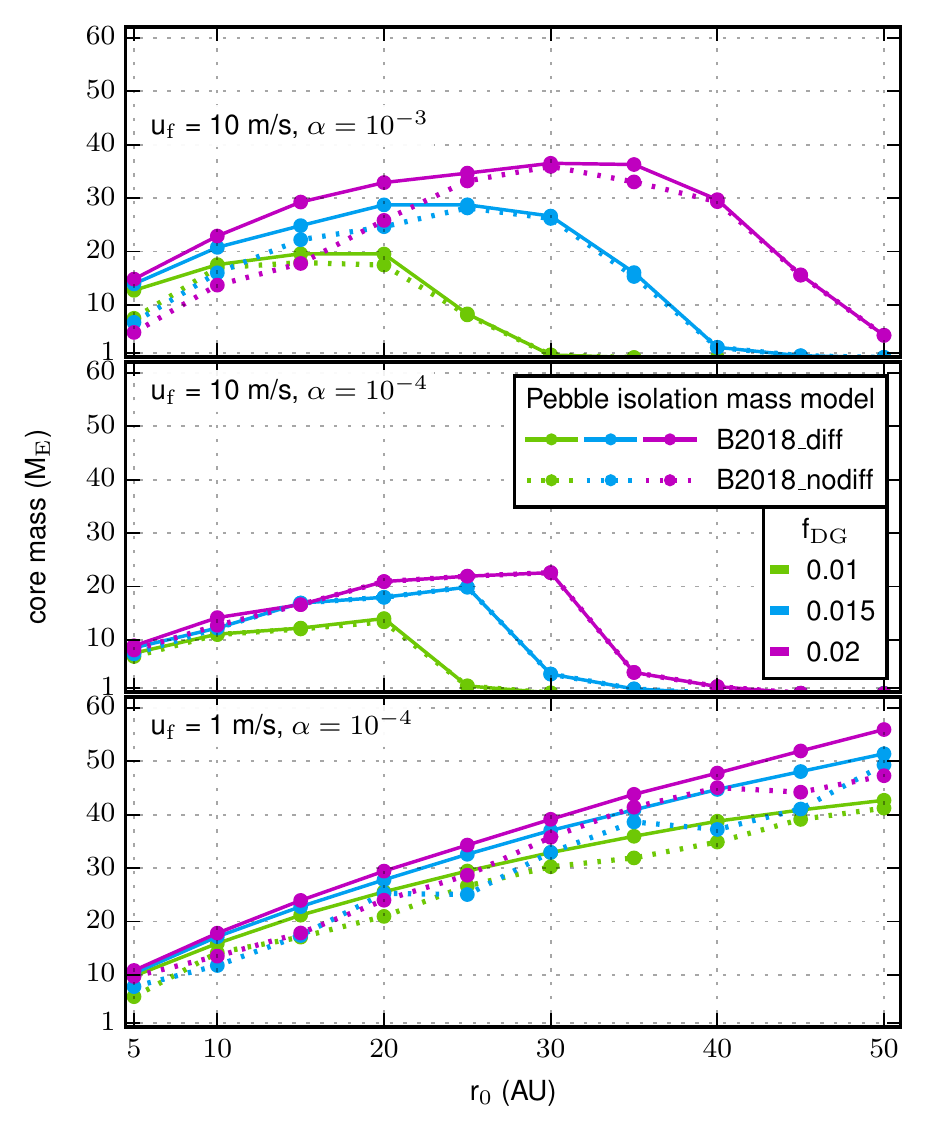} 
\caption{The dependence of core masses on turbulence strength and fragmentation velocity. B2018\_diff and B2018\_ndiff denote the  pebble isolation mass models of~\citetalias{bitsch2018}  with and without diffusion, respectively.}
\label{fig:fig08}
\end{figure}

The new formulation of pebble isolation mass in Equation~(\ref{eq:PIM}),  functionally depends on several parameters,  including the turbulence strength and pebble Stokes number; it relates with pebble Stokes number in such way that $M_{\rm{iso,~i}}\propto (1/\tau_{\rm{i}})$. As a consequence of Equation~(\ref{eq:PIM}), we assume in our model that different pebble species are sequentially isolated  by the growing planet when it starts to exert a pressure bump. This means that in our model, pebbles with large Stokes numbers are trapped exterior the planet’s orbit first, and those with small Stokes numbers are blocked last. We remind the reader that we did not model pressure bump due to the growing planet. This would require modelling a rather complex gap profile that is anchored self-consitently to viscous disc evolution as in~\cite{dullemond2018}, which is out of scope of this study and will be considered in future works.

As we illustrated in Figures~\ref{fig:fig003} and~\ref{fig:fig02}, pebbles may likely be blocked preferentially according to their sizes as a consequence of the pebble isolation prescription of \citetalias{bitsch2018}. For example, from both Figures~\ref{fig:fig003} and~\ref{fig:fig02}, all pebble species to the right of the vertical line are blocked at the planetary mass indicated on the line.

From the above notion, the planet then  reaches different isolation masses corresponding to each pebble species. At each pebble isolation stage, the planet continues to accrete smaller pebbles that are able to diffuse through the pressure bump that it generates, resulting in sustained core growth. However, when most of the pebbles have been blocked, the pebble accretion rate drops as the planet continues to slowly accrete the remaining small pebble species. For example, from Figure~\ref{fig:fig05}, for the parameter set of $u_{\rm{f}}=1~\rm{m/s}$ and $\alpha_{\rm{t}}=10^{-4}$, a planetary seed planted at 50 AU reaches its final core mass of $\sim 40~M_{\rm{E}}$ at around 20 AU, at which pebble isolation mass ranges from 33 -- 47 $M_{\rm{E}}$ for pebbles with Stokes number in the range 0.001 -- 0.1 (see Figure~~\ref{fig:fig02}). Before that, as seen in Figure~\ref{fig:fig05}, the accretion rate drops at around 30 AU without any significant increase in mass. This is because  from 30 AU down to 20 AU the planet is now slowly accreting small pebble species whose isolation mass is still high, while majority of the other pebble species have been blocked. Finally, the planet reaches an isolation mass that roughly corresponds to isolation mass of $\sim 40~M_{\rm{E}}$ for approximately 0.02 -- 0.1 cm pebble species at 20 AU as shown in Figure~\ref{fig:fig02}.

As shown in Figure~\ref{fig:fig02}, pebble isolation mass spans a wide range of values. We observe here that relatively large pebble species in the distribution need very similar pebble isolation mass in order to be blocked. On the other hand, the millimetre-sized pebbles  have a broader range of pebble isolation mass needed to block each species. This suggests that a planet may reach pebble isolation mass only once if it is accreting predominantly large-size pebble species. On the contrary, the planet may need to reach different isolation mass corresponding to each pebble species if the distribution is dominated by small-size pebble species.

The distributions in Figure~\ref{fig:fig02} further suggest that centimetre-sized pebbles produce relatively smaller cores since they are isolated at relatively low planetary masses compared with the smaller millimetre-sized pebbles. Thus, this suggests that building larger cores requires the presence of small-size pebbles, typically in the millimetre range, but more growth time will be needed to reach pebble isolation mass (see Figure~\ref{fig:fig04}). This would require either moderately turbulent discs with high fragmentation velocities or quiescent discs with low fragmentation velocities, both of which produce pebbles in millimetre-centimetre ranges. 

Furthermore, building massive cores requires that we implant planetary cores at far orbital distances as can be seen in Figure~\ref{fig:fig05}. This enables the planet to continue accreting before it reaches pebble isolation mass, which scales not only with pebble size but also orbital distance as in Figures~\ref{fig:fig02}. However, there appears a problem of fast planet migration that mostly drives the planet to the inner discs, unless type~I  migration rates are slowed via mechanisms like dynamical torques as in~\cite{ndugu2021}. As demonstrated here, fast core growth via multi-pebble accretion paradigm could also provide another pathway for the planet to reach high pebble isolation mass before it migrates rapidly to inner disc regions where the isolation mass is smaller. Such fast core growths can allow the planet to beat the rapid type I migration as suggested in~\cite{johansen2019}.

The combination of millimetre-centimetre size distribution thus has two main advantages: the centimetre-size species initially offer high accretion rates that enables the planet to rapidly become massive enough to accrete the smaller species more efficiently; the small species survive much longer in the disc and hence steady supply of material. This can then result in massive cores if multi-pebble accretion is taken into account together with diffusion as a determinant for pebble isolation mass.

The influence of turbulent diffusion in the context of multi-pebble accretion model is demonstrated in Figure~\ref{fig:fig08}. 
From the top panel of Figure~\ref{fig:fig08}, there is a significant difference in final core masses resulting from models with and without diffusion, especially in disc regions inside 20 AU. However, at wider initial growth locations, the planet has the same mass regardless of pebble isolation mass model used as shown in the top panel of Figure~\ref{fig:fig08}. In this case, the grain sizes may have similar isolation mass as previous discussed.

In the case of $u_{\mathrm{f}}=10~\mathrm{m/s}$  and $\alpha_{\mathrm{t}}=10^{-4}$, as shown in the middle panel of Figure~\ref{fig:fig08}, the difference between the pebble isolation mass schemes with and without diffusion is marginal. This is because, as already mentioned before, grain sizes are generally bigger at high fragmentation velocities, and with low turbulence, such large pebbles may not easily diffuse across pressure bump generated by the planet. This means the dependence of the final core mass on isolation mass for larger pebbles species may turn out to be inconsequential. 
However, for the same turbulence strength of $\alpha_{\mathrm{t}}=10^{-4}$ and a lower fragmentation velocity of for $u_{\mathrm{f}}=1~\mathrm{m/s}$, the effect of diffusion of pebble isolation mass is still manifested, as shown in the bottom panel of Figure~\ref{fig:fig08}. This is because low fragmentation velocities keep grain sizes small, which may be able to diffuse much more easily in less turbulent disc compared with the larger grains.

In consideration of sequential isolation of pebbles as discussed above, as more and more pebbles are trapped outside the planet’s orbit, the dust-to-gas ratio  and pebble flux would ultimately reduce in the inner disc. This would  affect growth of planetary cores interior to the planet’s orbit. In the outer orbit of the planet, the pile up of pebbles enhances dust-to-gas ratio at the pressure bump. This could trigger streaming instability and subsequent formation of planetesimals, thus reducing the amount of drifting pebbles, including those that would possibly diffuse through the pressure bump. This can then lead to reduced multi-pebble accretion rates and hence low mass cores.

If the idea that smaller pebbles may overcome the pressure bump holds, then their conversion into planetesimals at the pressure bump together with the other species would affect the accretion rate of the smaller pebbles. But the planetesimals formed may be potential targets for accretion by the planet. However, in our simulations we ignored the possible conversion of pebbles into planetesimals at the pressure bump, which may play a key role in the mass budget of the inward drifting pebbles. Here we instead focused on how the diffusion of smaller pebbles affects core growth rates.

In the light of the novel flow isolation mass~\citep{rosenthal2020}, we envisage that our results could substantially change. In particular, the flow isolation mass would limit the accretion of smaller, tightly coupled pebbles as they begin to interact with the planet’s atmosphere, and may simply flow past the planet without being accreted. Consequently, the final core masses would then be restricted to the isolation mass of the larger pebbles, which could be lower than the values reported here. Future comprehensive planet population synthesis studies should consider detailed comparison between the existing pebble isolation criterion and the flow mass paradigm in explaining the existing features of the observed gas giant planets. 

\subsection{Giant planet core formation}
Our results suggest that it is very challenging to form cores of giant planets when pebbles are much more coupled to the gas, for instance with $\tau_{\rm{i}}\lesssim0.001$, in agreement with~\citet{johansen2019}. Such pebbles tend to follow gas streamlines and interact gravitationally poorly with the protoplanets, making their capture difficult~\citep[see][for pebble capturing efficiencies by protoplanets]{guillot2014}. Furthermore, \citetalias{bitsch2018} have demonstrated that reaching pebble isolation mass for tightly coupled pebbles requires cores as massive as 50 $M_{\rm{E}}$. Reaching such a high core mass may prove difficult in the framework of dominant species, whose performance we have not rigorously tested here.

The Jupiter Near-polar Orbiter (JUNO) mission provided precise measurements of Jupiter's gravitational field, which has enabled better estimates of core masses. For instance, Jupiter's core mass was estimated to be 7 -- 25 $M_{\rm{E}}$~\citep{wahl2017} while ~\cite{debras2019} estimated 25 -- 30 $M_{\rm{E}}$ and 30 -- 45 $M_{\rm{E}}$ for non-compact and compact Jupiter cores, respectively. Other studies have inferred similar core masses for solar system gas giants: Jupiter (10 --40 $M_{\rm{E}}$~\citep{guillot1999} and 37 $M_{\rm{E}}$~\citep{thorngren2016}); Saturn (20 --30 $M_{\rm{E}}$~\citep{guillot1999} and 27 $M_{\rm{E}}$~\citep{thorngren2016}). Many models have considered giant impacts to account for Jupiter's highly enriched core with heavy elements~\citep[e.g.,][]{liu2019,ginzburg2020}. We have tested that it is quite challenging to obtain such high core masses using the classical pebble accretion model adopted in this work, unless we at least unphysically increase pebble concentration and consider discs with longer life times. Moreover, the cores of giant planets that migrated to $<10$ AU could have started their growth beyond between 15 -- 30 AU~\citep{johansen2019}. In this orbital domain, we found it difficult to grow cores to masses above 25 -- 30 $M_{\rm{E}}$ without evoking accretion of multiple pebble species. We therefore think that through our multi-pebble accretion scheme, massive cores of gas planets that match the results the JUNO mission can form.

\section{Conclusion}\label{conclusions}
In this work, we have extended the pebble accretion paradigm, where core growth proceeds by concurrent accretion of multiple pebble species. We took into account the dependence of pebble isolation mass on turbulence parameter and pebble size. In our model, the final planetary core masses were set by the planet mass needed to block the smallest sized pebble species in the distribution of grain sizes under consideration. We self-consistently reconstructed grain distribution throughout the core growth process using the reconstruction model of~\cite{birnstiel2015}. For the dust evolution, we employed the two-population model of~\cite{birnstiel2012}.

Our simulations yielded a diversity of planetary cores with a wide range of core masses. The final outcome of core masses is primarily dictated by a combination of dust-to-gas ratio, fragmentation velocities and turbulence strength that underpin the size distributions. Under a suitable set of disc conditions, cores of gas giants can form at orbits as far out as 50 AU.

Our work has several other important consequences. Firstly, recent studies~\citep{wahl2017,debras2019} modelled Jupiter's core to contain heavy elements totalling to 25 -- 45 $M_{\rm{E}}$, adding a further constraint on core accretion models. It is a difficult to invoke pebble accretion to explain such massive core build up. Moreover, within the limitations of our numerical simulations, we could not easily build massive cores well above $\sim30~M_{\rm{E}}$ using the classical scenario of dominant species. If we are to invoke pebble accretion to explain assembly of such massive cores, then our model of concurrent accretion of multi-pebble species would provide a possible mechanism to explain this enigma.

Secondly, in the study of metallicity correlation of extrasolar planets,~\citet{guillot2006b} found that their sample planets contained $\sim$ 20 -- 100 $M_{\rm{E}}$ in heavy elements. The work of~\cite{thorngren2016} also suggests a strong correlation between the planetary mass and amount of heavy elements accreted by the planet. If these results are confirmed, then the accumulation of such large amount of heavy material is in conflict with classical core accretion model of planet formation. However, some recent solutions have been proposed, for example, the merger of giant planets via collisions~\citep{liu2019, ginzburg2020} or the accretion of the evaporated material may increase the heavy element content of the giant planets~\citep{schneider2021}.

~Another important aspect of classical core accretion that we did not account for in this study is the accretion of planetesimals. Although gas giant planet cores form with difficulties via planetesimal accretion, the capture of planetesimals by the planet can also enrich the planet with heavy elements~\citep[][]{shibata2019,shibata2020,venturini2020a}. Nevertheless, as we have demonstrated here, it is possible that these planets consumed several different dust species if their formation proceeded through pebble accretion paradigm, and can form such massive gas giant planet cores even at the disc's outskirts.  

Thirdly, a diversity of gaps and rings over a wide orbital regime has been discovered in many discs through the ALMA program~\citep[e.g.,][]{huang2018,huang2018b,huang2018c,long2018} and the DSHARP campaign~\citep[e.g.,][]{andrews2018}. Planet-disc interactions have been evoked as one possible way to explain the occurrence of these structures, though this is subject to discussion~\citep{ndugu2019}. Recently, in a sample of 16 ALMA discs studied by~\citet{vandermarel2019}, the authors found some substructures at orbital distances far beyond 30 AU, and found evidence for presence of massive planets at those orbital distances. An important question as to whether or not these substructures are indeed caused by growing planets has been put forth~\citep[e.g.,][]{lodato2019, ndugu2019}. If it is true that planets may be responsible for opening gaps at such wider orbital distances, then the existing core accretion models are missing important ingredients for explaining the formation of planets at these remote locations. It is, however, possible to form gas giant planets at these wide orbits if: (i) concurrent accretion of multiple pebble species is accounted for, (ii) it is assumed that the giant planets at such wide orbits form via gravitational instability paradigm~\citep[][]{boss1997,boley2009,armitage2010} and (iii) assuming the observed rings are caused by other phenomena such as MRI other than an on-going planet formation. The rings could in turn be hot spots for planet formation at such wide orbits~\citep{morbidelli2020}. Thus the precise explanation for the origin of the observed gaps/rings at wider orbits is a complicated one, more like a {\it Chicken-Egg} problem.

 Although our model has shed some light on the formation of massive giant planets at wider orbits of discs, it did not capture important planet formation aspects like competition of multiple cores for the available building block and orbital manipulation by N-body interactions between cores. We therefore recommend future models that study  multi-pebble species accretion to consider core growth competition and the N-body paradigm for detailed and substantial explanation of the formation of gas giant planets at wider orbits.

\section*{Acknowledgments}
We thank the anonymous referee for the insightful reports that helped us to improve this paper. G.A. further thanks Michiel Lambrechts for constructive and inspirational comments. 
We thank Swedish International Development Cooperation Agency (SIDA) for financial support through International Science Programme (ISP) Uppsala University Sweden to the East Africa Astronomical Research Network (EAARN). 
\section*{Data availability}
For the purpose of reproducibility, the code used to obtain results in this paper will be provided upon reasonable request.


\bibliography{./Biblio.bib}
\bibliographystyle{mnras}




\appendix
\section{The dust distributions and mass fractions}\label{sec:appendixA}
\begin{figure*}
\includegraphics[width=\textwidth]{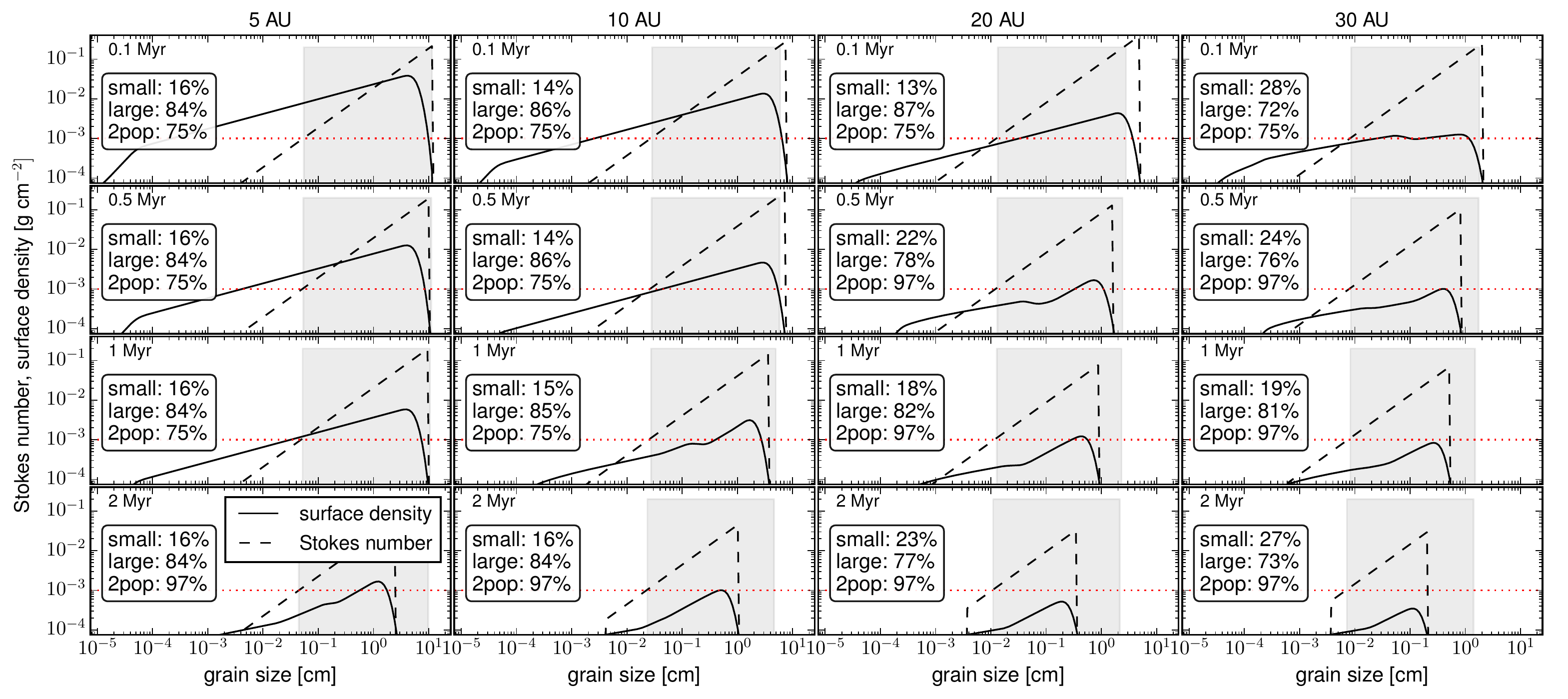} 
\caption{The grain surface density and Stokes number as a function of grain size reconstructed from the two-population model at 5 AU, 10 AU, 20 AU and 50 AU at time 0.1 Myr, 0.5 Myr, 1 Myr and 2 Myr. On the plots, we show the fractions of {\it{small}} and {\it{large}} grains that have Stokes numbers less and greater than 0.001, respectively. The shaded regions show large grains with Stokes number 0.001, that were mostly accreted in our growth model. Here, {\it{2pop}} is the fraction of the large dust population from the two-population simulations that contains most of the solid mass. The dust evolution was performed with $u_{\mathrm{f}}=10 \mathrm{m/s}$, $\alpha_{\mathrm{t}}=10^{-3}$ and $f_{\mathrm{DG}}=0.01$ for 2 Myrs.}
\label{fig:figappendix01}
\end{figure*}
\begin{figure*}
\includegraphics[width=0.99\textwidth]{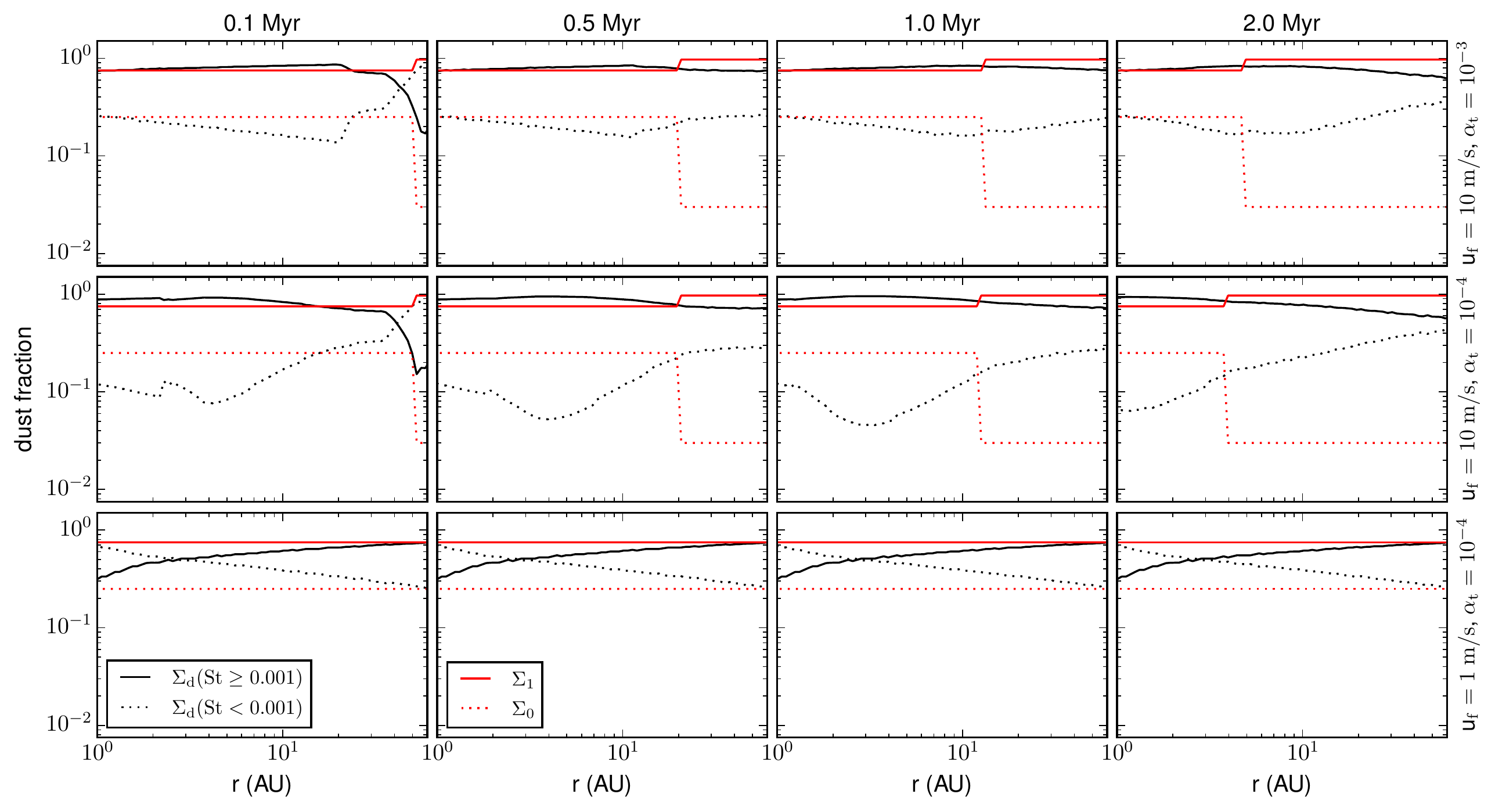} 
\caption{The radial mass fractions of {\it{small}} (with Stokes number < 0.001) and {\it{large}} ( with Stokes number $\ge 0.001$) grains. $\Sigma_0$ and $\Sigma_1$ are the small and large dust populations from the two-dust population model. Each row shows simulations with different combinations of fragmentation velocity and turbulence strength, and initial dust-to-gas ratio of $f_{\mathrm{DG}}=0.01$.
\label{fig:figappendix02}}
\end{figure*}

In Figure~\ref{fig:figappendix01}, we show the variation of pebble surface density and the Stokes number as a function of grain for the dust size distribution reconstructed at radial distances 5 AU, 10 AU, 20 AU and 50 AU, for $\alpha_{\mathrm{t}}=10^{-3}$, u$_{\mathrm{f}}=10~\mathrm{m/s} $ and f$_\mathrm{DG}=0.01$. On each plot, we show the fractions of pebbles that have Stokes number greater and less than 0.001. Grains with Stokes number $\ge 0.001$ are shown by the shaded region. We also show the fractions of the large and small populations from the two-population model. In the two-population scheme, grain sizes are calculated and fixed by growth, fragmentation and drift limits. Here, in the inner and outer disc regions, the fragmentation and drift limits dominate, respectively with 0.75 and 0.97 of the mass constituting the large population. As shown in the plots, the reconstructed grains distributions with Stokes number $\ge 0.001$ dominate in the inner disc regions, $\le$ 30 AU, which carry over 80\% of the mass, similar to the fractions of the two dust population model.

Furthermore, as shown in Figure~\ref{fig:figappendix01}, very few grains have Stokes number greater 0.1, and most of the mass in reconstructed grain distribution is carried by the population of grains with Stokes number $\ge 0.001$. In addition, the accretion of pebbles with Stokes number < 0.001 is generally inefficient since these pebbles are usually well coupled to the gas~\citep{guillot2014}, and we observed the same trend in our simulations.  Thus in our simulations, the embryos mostly accrete pebbles with Stokes number $\ge 0.001$.

In Figure~\ref{fig:figappendix02}, we show dust fractions as a function of radial distance for the different combinations of fragmentation velocities and turbulence strengths, and for both the simulated two-population and the reconstructed grain size distributions. In the simulations with fragmentation velocity of 10 m/s, most of the mass is carried by grains with Stokes number $\ge 0.001$ within 50 AU as shown by the panels in the first and second rows of Figure~\ref{fig:figappendix02}.  On the other hand, for the case of fragmentation velocity of 1 m/s, the fraction of pebbles with Stokes number $\ge 0.001$ begins to approach that of the large population simulated using the two dust model at orbital distances greater than 5 AU. Here, within 5 AU, pebbles with Stokes number less than 0.001 dominate. This may suggest a stronger fragmentation inside 5 AU that produces smaller grains, which then have small Stokes numbers compared with other parts of the disc.

\section{The recipe for concurrent accretion of multi-pebble species}\label{sec:appendixB}
In this Section we give a brief description of the recipe that we used in our numerical simulations involving concurrent accretion of multi-pebble species in reference to the full-size distribution as in~\citet{birnstiel2012}. It is a simple extension of the implementation of single species model that can readily be adopted for other models of particle distribution.
At each time snapshot and orbital distance referenced by j and k respectively, we implement core accretion of multiple pebble species as follows:
\begin{enumerate}[Step 1.]
\item Initialise planet mass, $M_{\rm{j=0,k=0}}$, orbital distance, $r_{\rm{k=0}}$.
\item Calculate the disc parameters at r$_{\rm{k}}$ and time t$_{\rm{j}}$: \label{Step2}
\item Obtain the logarithmic distribution of N pebble sizes $R_{\rm{i+1}}=1.12R_{\rm{i}}$  as in~\cite{birnstiel2011}.\label{Step3}
\item From the logarithmic particle size distribution, calculate the pebble surface density, $\Sigma_{\rm{peb, i}}$,  for each of the N species according the reconstruction scheme in~\citet{birnstiel2015} and calculate the Stokes number, $\tau_{\rm{i}}$, for each species in Epstein regime as
\begin{equation}
\tau_{\rm{i}}= \frac{\rho_{\bullet}\pi R_{\rm{i}}}{2\Sigma_{g}}.
\end{equation}\label{Step4}
\item Calculate pebble isolation mass for each species according to Equation~\ref{eq:PIM}.\label{Step5}
\item If the planet mass is larger than the pebble isolation mass of pebble species with the smallest Stokes number in the size distribution, stop time integration. This is because as the planet grows, pebbles with larger Stokes numbers are isolated first, while the ones with smaller Stokes numbers are isolated last, depending on the mass of the planet (see Equation~(\ref{eq:PIM}) and more explanation in \citetalias{bitsch2018}). Hence the core mass is determined by the isolation mass of pebbles with smallest Stokes numbers.\label{Step6}
\item If planet mass is larger than the isolation mass of the i-th species, stop accretion of the i-th species. Otherwise accrete the i-th species by calculating the core accretion rate $\dot{M}_{\rm{core,i}}$ using Equations~(\ref{eq:06}), (\ref{eq:07}) and (\ref{eq:CoreGrowthRegime}).\label{Step7}
\item Calculate the total core growth rate by consolidating contributions from each species:
 \begin{equation}
  \dot{M}_{\rm{j,k}}=\sum_{\rm{i}} \dot{M}_{\rm{core,i}}.
 \end{equation}\label{Step8}
\item Calculate the new planet mass from
 \begin{equation}
  M_{\rm{j+1,k}}=M_{\rm{j,k}}+\dot{M}_{\rm{j,k}}\delta t,
 \end{equation}\label{Step9}
 where $\delta t$ is the timestep.
\item Calculate the migration rate and the new orbital location of the growing core.\label{Step10}
\item Increase time by $\delta t$ and repeat from Step~\ref{Step2} until the condition in Step~\ref{Step6} is fulfilled.\label{Step11}
\end{enumerate}

\bsp 
\label{lastpage}
\end{document}